\documentclass[reprint, amsmath,amssymb,aps]{revtex4-2}

\usepackage{graphicx}
\usepackage{dcolumn}
\usepackage{bm}
\usepackage{xcolor}
\usepackage{color}

\begin{document}

\preprint{APS/123-QED}

\title{Perturbation of charge density waves in $1T$-TiSe$_2$}

\author{Imrankhan Mulani}
\author{Umashankar Rajput}
\author{Luminita Harnagea}
\author{Aparna Deshpande}
\affiliation{Indian Institute of Science Education and Research (IISER) Pune,Dr. Homi Bhabha Road, Pashan, 
Pune 411008, India }

\date{\today}% It is always \today, today,

\begin{abstract}
In this study, using low-temperature scanning tunneling microscopy (STM), we focus on understanding the native defects in pristine  $1T$-TiSe$_2$ at the atomic scale. We probe how they perturb the charge density waves (CDWs) and lead to local CDW modulated region formation. These defects influence the correlation length of CDWs. We establish a connection between suppression of CDWs, Ti intercalation, and show how this supports the exciton condensation model of CDW formation in $1T$-TiSe$_2$. 

\end{abstract}

%\tableofcontents
\maketitle

\section{Introduction}

Transition metal dichalcogenides (TMDCs) are versatile materials that exhibit phenomena like strongly correlated phases to multifaceted tunable properties for applications in flexible electronics, optoelectronics, spintronics, and energy harvesting \cite{Tedstone2016, mueller2018exciton, pallecchi2020review, huang2020recent}. Many exotic physical phenomena, like spin-valley interaction, exciton-polariton states, charge density wave (CDW) states, are observed in the TMDCs. TMDCs have a characteristic layered structure, tunable bandgap, and strong spin-orbit coupling favoring such applications. The chemical composition of TMDCs is of the form $MX_2$ where $M$ denotes transition metal (like $Ti$, $Mo$, $W$) and $X$ represents chalcogen (like $S$, $Se$, and $Te$). Despite having a similar structure, these materials can be categorized as insulating, semiconducting, semi-metallic, metallic, or superconducting, depending on their chemical composition. The non-bonding $d$ band and the extent of its electron filling endow these compounds with diverse properties. Bulk crystals of TMDCs consist of two-dimensional (2D) layers bonded by weak van der Waals interactions, rendering them easy to exfoliate. Depending on the arrangements of atoms, structural polytypes of 2D TMDCs can be categorized as $1T$ (tetragonal symmetry with octahedral coordination), $2H$ (hexagonal symmetry with trigonal prismatic coordination), and $3R$ (rhombohedral symmetry with trigonal prismatic coordination). The layer degree of freedom provides an additional parameter to tune the properties of TMDCs \cite{book1}.

One of the group IV TMDCs hosting interesting properties is $1T$-TiSe$_2$. It is a semimetal with a small indirect bandgap \cite{isomaki1981gaps}. At temperature $T_{CDW}\sim200K$ $1T$-TiSe$_2$ undergoes a second-order phase transition to the charge density wave (CDW) phase \cite{di1976electronic}. Charge density wave is a many-body, correlated electrons phenomenon, where a periodic distortion modulates electron density. CDWs, depending on the ratio of their wavelength and lattice parameters, are classified as commensurate (CCDW), nearly commensurate (NCCDW), or incommensurate (ICDW) \cite{gruner1988dynamics,rossnagel2011origin}. Pristine $1T$-TiSe$_2$ shows CCDW order below $200K$ \cite{di1976electronic}. Charge carrier doping via electrochemical ionic gating in $1T$-TiSe$_2$ leads to suppression of CCDW, emergence of ICDW and superconducting phase  at lower temperature \cite{li2016controlling}. The CDW transition temperature drops significantly for $1T$-TiSe$_2$ doped with copper. Superconductivity emerges in Cu$_x$TiSe$_2$ for $x\sim0.04$, maxima of superconducting transition temperature reaches at $x\sim0.08$ \cite{morosan2006superconductivity}. ICDW phase coexists with superconducting phase under applied pressure \cite{joe2014emergence} and copper doping \cite{kogar2017observation}. The mechanism of formation of CDWs in $1T$-TiSe$_2$ is still a matter of debate. The two favored mechanisms are excitonic insulator phase \cite{jerome1967excitonic} and band type Jahn-Teller effect \cite{hughes1977structural}. Excitons are bound states of electrons and holes. Excitons are bosons and can form a Bose-Einstein condensate \cite{kunevs2015excitonic,seki2014excitonic,brinkman2005electron,wang2019evidence}. The excitonic insulator phase can arise in a small gap semiconductor or semimetal, where excitons form spontaneously because of low carrier density to screen the attractive Coulomb interaction between electrons and holes.  In the Jahn-Teller effect, the lattice spontaneously distorts to lift the degeneracy and reach the lower symmetry state because of the interaction between phonons and degenerate electron states. Jahn-Teller like mechanism is independent of free carrier concentration.

$1T$-TiSe$_2$ is a nonstoichiometric compound. The concentration of $Ti$ in a crystal depends on the temperature at which the crystal is grown. Commonly used method for TMDC crystal growth is chemical vapour transport (CVT). Higher growth temperature generates higher chalcogen pressure resulting in a crystal with intercalation of metal \cite{shkvarin2020effect}. An indirect inference of presence of excess $Ti$ can be carried out through temperature-dependent resistivity measurements. $1T$-TiSe$_2$ shows anomalous resistivity peak near CDW phase transition. As the growth temperature increases, the crystal becomes more metallic due to excess $Ti$ and hence suppresses the anomalous resistivity peak \cite{di1976electronic,krasavin1998effect}. $1T$-TiSe$_2$ shows a small indirect band gap above the transition temperature $T_{CDW}\sim200K$ \cite{watson2019orbital,chen2015charge}. Angle resolved photo emission spectroscopy (ARPES) experiments show a small band gap below CDW transition \cite{rossnagel2002charge}. Recent experimental evidence suggests excitonic insulator mechanism \cite{cercellier2007evidence,hildebrand2016short,monney2009spontaneous,monney2011exciton,cazzaniga2012ab,kogar2017signatures}. Theoretical investigations suggest that excitonic condensation can be either superfluid \cite{snoke2002spontaneous} or an insulator \cite{kohn1970two}. Chiral nature of CDWs in pristine and Cu-doped $1T$-TiSe$_2$ has been reported using polarized optical reflectometry and STM \cite{ishioka2010chiral,ishioka2011charge,iavarone2012evolution}. Suppression of CDW results in superconducting phase transition with application of pressure\cite{kusmartseva2009pressure} via electrostatic gating \cite{li2016controlling} or with Cu intercalation \cite{morosan2006superconductivity} or Pd intercalation \cite{morosan2010multiple}. $1T$-TiSe$_2$ becomes insulating after Pt doping\cite{chen2015chemical}. This implies that doping affects the nature of CDWs and electronic properties of $1T$-TiSe$_2$.

A correlated phenomenon, like CDWs, is sensitive to the presence of defects. Here, using low-temperature STM, we focus on the behavior of CDWs at the atomic scale in the presence of intrinsic defects. We investigated the relationship between growth conditions and intrinsic defects in crystals grown by different synthesis routes. We referred to them as sample A (crystal was grown by chemical vapor transport (CVT) using iodine (I$_2$)) and sample B (crystal was grown using $Se$ as self-flux) in the subsequent discussion.

\section{Methods}

The experiments were performed by using an Omicron ultra-high vacuum (UHV) low temperature(LT) scanning tunneling microscope (STM) at $77K$ with a base pressure of  5$\cdot10^{–11}$  mbar. Two $1T$-TiSe$_2$ crystals were studied.  Sample A was grown using CVT with $I_2$ as the transport agent. Sample B synthesized using the flux zone method by 2D semiconductors, USA \cite{2Dsemi}. For STM studies of both the samples, $1T$-TiSe$_2$ single crystal surface was prepared by cleaving it in the UHV sample preparation chamber. The surface quality was checked with STM at $77K$. A clean surface at the atomic level was inferred from several STM images of different scan sizes. Electrochemically etched tungsten($W$) tips used for imaging. Several trials of imaging were performed with different tips. Topography measurements were taken in constant current mode. In our set up bias voltage is applied to the sample. The images were processed using the image analysis software SPIP 6.0.9 (Image Metrology). MATLAB (by The MathWorks Inc.) was used for numerical calculations.

\section{Results and discussion}
$1T$-TiSe$_2$ crystal is a stack of monolayers bonded by van der Waals interaction. Each monolayer consists of titanium ($Ti$) atoms sandwiched between two sheets of selenium ($Se$) atoms (Fig.1\textbf{(a)}). Each $Ti$ atom is surrounded by six $Se$ atoms in octahedral coordination, as shown in Fig.1\textbf{(a)}. The space group of $1T$-TiSe$_2$ is $P\overline3m1(no.164)$. The crystal cleaves along $(001)$ planes were exposing $Se$ atoms in a hexagonal symmetric pattern. In constant current topography images, STM tip probes the topmost $Se$ atoms \cite{slough1988scanning,hildebrand2014doping}. The tunneling current is mainly due to the contribution from the $4p$ orbital of $Se$ atoms. The preferred site for $Ti$ atom intercalation is schematically shown in Fig.1\textbf{(b)}, whereas $O$ substitution site shown in Fig.1\textbf{(c)}. Iodine is used as a transport agent in CVT method of crystal growth. Substitution of $Se$ atom by $I$ is another defect commonly found in CVT grown samples. $I$ substitution occupies same sites as $O$ substitution. Defects in the STM topography image of sample A in Fig.1\textbf{(d)} and sample B Fig.1\textbf{(e)} show up as localized bright regions. Note that the defects are less in sample B than in sample A.

\begin{figure}[h]
\includegraphics[width=8.6cm]{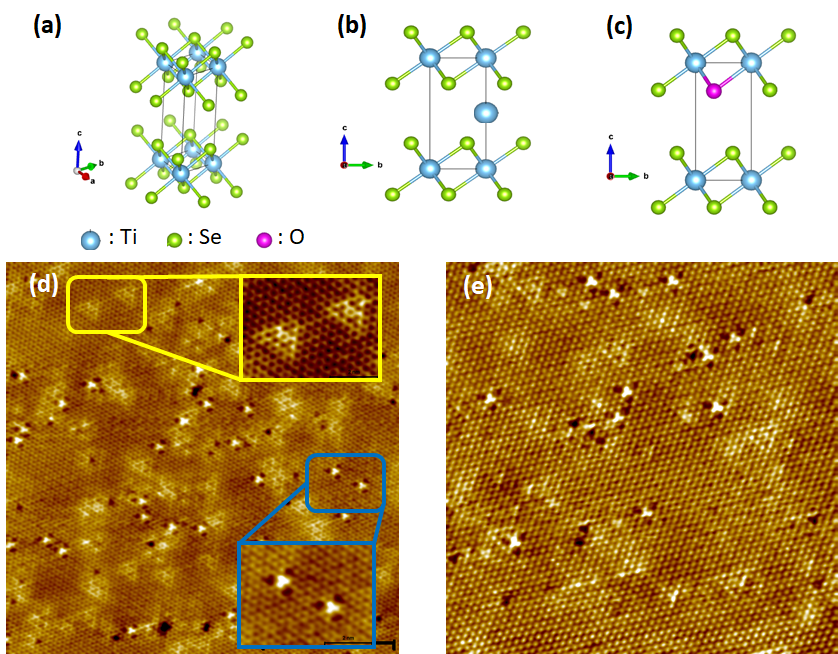}% Here is how to import EPS art
\caption{ \textbf{(a)} Schematics of $1T$-TiSe$_2$ unit cell viewed from an angle, $Ti$ atoms in octahedral coordination, $Se$ atoms form a triangular lattice. \textbf{(b)} Schematics of $Ti$ intercalation site \textbf{(c)} Schematic of $O$ substitution site \textbf{(d)} Empty-states constant current STM image of $30 nm \times 30 nm$ area of sample A, Enlarged view of $Ti$ intercalation sites are in yellow rounded rectangle inset and that of $O$ substitution sites are in blue rounded rectangle inset. $V_{bias} = 300$mV and $I_{set} = 300$pA. \textbf{(e)} Empty-states constant current STM image of $20 nm \times20 nm$ area of sample B ,$V_{bias} = 300$mV and $I_{set} = 500$pA.}
\end{figure}

In Fig.1\textbf{(d)} the inset in yellow rounded rectangle shows the enlarged atomic-scale image of the defect, which can be identified as $Ti$ intercalation site. Intercalated atoms occupy octahedral sites or tetrahedral sites in the van der Waals gap. For $Ti$, favorable sites are octahedral sites because of their more significant coordination number, and spacing between two nearby octahedral sites is more considerable than tetrahedral sites \cite{fan2017theoretical}. The characteristic appearance of intercalated $Ti$ sites can be attributed to change in the local density of states \cite{may2011influence}. These defects appear bright in filled-states STM images (at negative bias voltages). 

There were defects with a strongly localized density of states at the topmost $Se$ atoms surrounded by depleted region in empty-states STM images (at positive bias voltages). Those defects can be identified with oxygen ($O$) substitution of the $Se$ atom in the lower layer. $I$ substitution sites are hard to recognize as they appear very faint compare to $O$ sites \cite{hildebrand2014doping}. Substitution of $Se$ by $O$ or $I$ on topmost layer is not favored energetically \cite{hildebrand2014doping}. Oxygen, present as an impurity in iodine and selenium precursor during growth, acts as a native dopant. Sample A has been observed to possess more $Ti$ intercalation sites than sample B (Fig.1\textbf{(d)} and 1\textbf{(e)}). Using STM images taken in different regions of both samples, the average number of intercalation sites were counted. Employing this method, we found that the average count of $Ti$ intercalation sites per unit cell for sample A is $0.6\%$ as compared to $0.2\%$ sites per unit cell for sample B. This observation supports the dependence of $Ti$ intercalation sites' concentration on sample growth conditions \cite {di1976electronic}.

\begin{figure}[h]
\includegraphics[width=8cm]{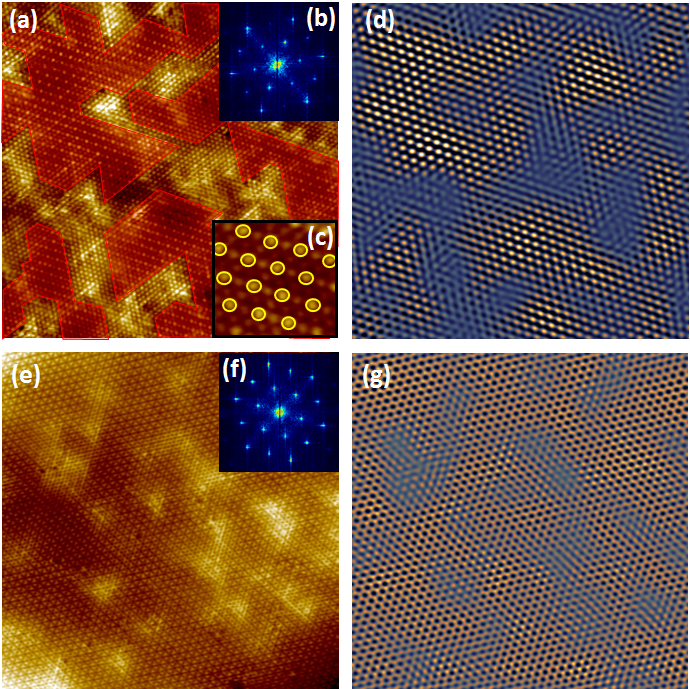}% Here is how to import EPS art
\caption{ \textbf{(a)} Filled-states constant current STM image of 30 nm X 30 nm area of sample A grown by CVT method, $V_{bias} = - 400$ mV $I_{set} = 300$ pA.Areas marked by red are CDW modulated regions, separated by non-modulated regions induced by Ti intercalation sites. \textbf{(b)} Fourier transform of topography image in \textbf{(a)}. \textbf{(c)} An enlarged image showing 2 X 2 superstructure with CDW peaks marked by yellow circles \textbf{(d)} Selective inverse Fourier transform of CDW peaks in \textbf{(b)}, \textbf{(e)} Empty-states constant current STM image of 30 nm X 30 nm area of sample B grown by flux zone method, $V_{bias} = - 300$mV $I_{set} = 400$pA, \textbf{(f)} Fourier transform of topography image in \textbf{(e)}. \textbf{(g)} Selective inverse Fourier transform of CDW peaks in \textbf{(f)}}
\end{figure}

Intercalation and substitution change the material's carrier concentration, leading to a change in the material's electronic properties. Correlated states like CDWs are sensitive to external factors like mechanical stress, external pressure, and carrier concentration. Point defects can affect CDWs by changing carrier density locally and introducing localized distortions in lattice \cite{shkvarin2020effect}. STM topography images were obtained at different bias voltages, spatially resolving the CDWs. Fig.2\textbf{(a)} and 2\textbf{(e)} shows STM images of sample A and B with different $Ti$ intercalation and $O$ substitution densities. Real space topography shows the clear $2\times2$ superlattice of CDW. Fig.2\textbf{(c)} is an enlarged image with atomic resolution showing CDWs forming superstructure (marked by yellow circles). Fourier transform (FT) of topography images (Fig.2\textbf{(b)} and 2\textbf{(f)}) yields contributions of lattice and CDW superlattice as six-fold symmetry peaks. Peaks in FT images affirm that the wavelength of CDW is twice the lattice spacing (first-order Bragg peak). To reduce background noise and understand the spatial variation of CDW amplitude, we performed an inverse transform of CDW FT peaks. Inverse Fourier transform (IFT) image clearly shows diminishing CDW amplitudes in the presence of $Ti$ defects in Fig.2\textbf{(d)} and 2\textbf{(g)}. In the presence of a large number of defects, CDW modulated regions can be easily visually identified from IFT images (Fig.2\textbf{(d)}). In $1T$-TiSe$_2$ with more $Ti$ intercalation density, domains of CDWs are formed by the presence of defects. 2 X 2 superstructure of CDWs changes to 1 X 1 modulation near intercalation sites, which forms the boundaries of CDW modulated region (Fig.2\textbf{(a)} and 2\textbf{(c)}). Also, the IFT image is used as a guide to locate these boundaries. Different CDW modulated regions are marked in red in Fig.2\textbf{(a)}. CDW modulated regions separated by boundaries was not observed in Fig.2\textbf{(e)} containing fewer intercalation sites. IFT in Fig.\textbf{(g)} shows the reduced amplitude of CDWs in the presence of intercalation sites but no formation of separate regions. $1T$-TiSe$_2$ thin flakes are reported to oxidise quickly in the atmospheric conditions and disrupt the CDW phase \cite{sun2017suppression}. The oxidation of $1T$-TiSe$_2$ was minimised by cleaving the crystals in our UHV sample preparation chamber(refer to section II. METHODS). $O$ and $I$ substitution sites increases the density of states in the unoccupied states away from the Fermi energy \cite{hildebrand2014doping}. The CDW gap near Fermi level remains unaffected by $O$ or $I$ substitution.

\begin{figure}[h]
\includegraphics[width=7cm]{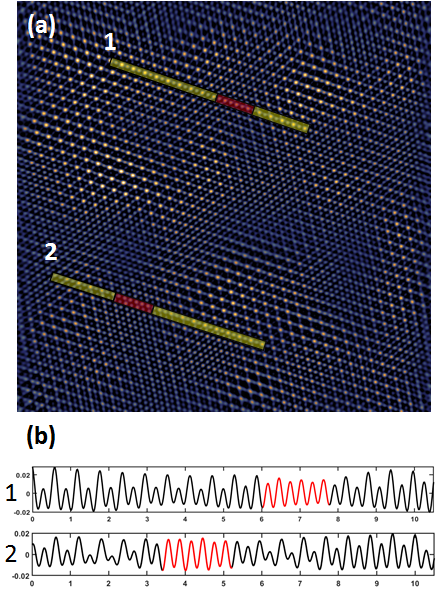}
\caption{\textbf{(a)} Inverse Fourier transform image of Bragg peaks and CDW peaks of STM image shown in Fig.2\textbf{(a)} of sample A. Line sections 1 and 2 passing through domain boundary (marked by red region) plotted in \textbf{(b)} Red part of line section indicates no CDW modulation. No phase shift observed across the domain wall.}
\end{figure}

Metal intercalation of $1T$-TiSe$_2$ at certain doping concentration shows phase shift of CDW modulation across the domain walls \cite{hildebrand2016short,jaouen2019phase}.$Cu$ intercalation in $1T$-TiSe$_2$ results in incommensurate CDW phase with localized commensurate CDWs separated by domain walls at critical doping concentration \cite{yan2017influence}. $Ti$, like $Cu$, is electropositive and suppresses the CDW phase. $Cu$ intercalation into van der Waals gap of $1T$-TiSe$_2$ was stabilized by transfer of charge from $Cu$ to $Se$ atoms in the adjacent layers \cite{jishi2008electronic}. Similarly, the $Ti$ intercalant gets stabilized by charge transfer.  A significant overlap between the atomic orbitals of intercalated $Ti$ and $Se$ atoms in the layer results in the intercalant sites' peculiar appearance in the STM images. To better understand phase shift of CDWs across domain boundaries, IFT of both lattice Bragg peaks and CDW peaks of Fig.2\textbf{(a)} were obtained, as shown in Fig.3\textbf{(a)}. A line section of electron density indicates the suppression of CDW modulation at the boundary (Fig.3\textbf{(b)}).

Interestingly we observed no phase shift in the CDW modulation across the boundaries.  We can infer that the increased concentration of $Ti$ intercalation leads to the disruption of CDW and reduces long-range order to small regions \cite{novello2015scanning}.

The electropositive nature of $Ti$ intercalation defect and formation of CDW modulated regions and non-modulated boundaries in excess $Ti$ intercalation may provide some insight into the mechanism of formation of CDWs. Considering the results as mentioned above, we offer a hypothesis for suppression of CDWs by $Ti$ intercalation. Semimetals or small gap semiconductors at lower temperatures can undergo CDW phase transition by excitonic condensation mechanism. Exciton is a bound state of electrons and holes, mediated by Coulomb interaction, a composite boson. For $1T$-TiSe$_2$ a small carrier density and a weakly screened Coulomb interaction can result in exciton formation. Excitons are formed from electrons in the pocket at the $L$ point in the Brillouin zone (BZ) and holes from the pocket at $\Gamma$ point in the energy band stucture of $1T$-TiSe$_2$. Being a composite boson, it undergoes condensation at $T_{CDW}\sim200 K$ as the number of excitons increases. Excitonic condensation leads to a periodic modulation of charge density, a CDW state. The extent to which CDWs are modulated in $1T$-TiSe$_2$ is intercalation dependent. 
If we assume excitonic condensation as the mechanism of CDW state in $1T$-TiSe$_2$ then the suppression of CDWs and formation of domain boundaries can be explained by the electropositive nature of $Ti$ intercalation.

The first-principles energy band calculations show that the electron pocket at $L$ point in BZ is derived mainly from $Ti$ $3d$ states. Hole pockets at $\Gamma$ are derived from $Se$ $2p$ states hybridized with $Ti$ $3d$ states \cite{zunger1978band}. At Fermi energy, more than half the density of states (DOS) results from $Ti$ states, and the rest comes from $Se$ \cite{jishi2008electronic}. As $Ti$ intercalation's character is electropositive, the addition of $Ti$ results in the acquisition of electrons. Because of $Ti$'s larger DOS, electrons added to the system go to $Ti$ derived states around the Fermi level.$Ti$ intercalation now causes the enhancement of electronic states at Fermi energy, resulting in the strong screening of Coulomb interaction between electrons and holes. This screening results in the reduction of excitonic pair and suppression of charge density waves.

\begin{figure}[h]
\includegraphics[width=8.5cm]{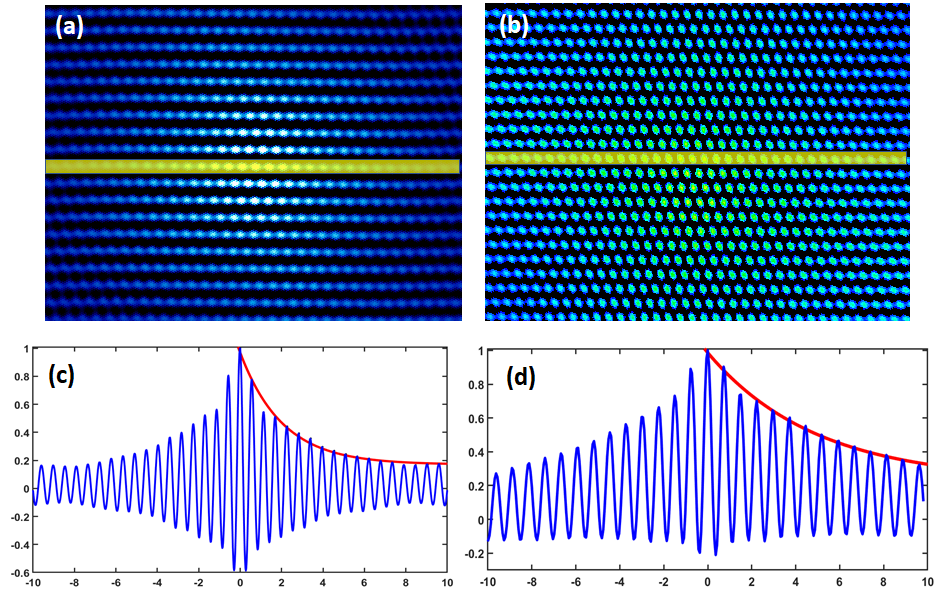}
\caption{ \textbf{(a)} Autocorrelation of 30 nm x 30 nm area of sample A , \textbf{(b)} Autocorrelation of 30 nm x 30 nm area of sample B, \textbf{(c)} and \textbf{(d)} Line section of autocorrelation of sample A shown in \textbf{(a)} and autocorrelation of sample B shown in \textbf{(b)} along the CDW direction respectively, red curve indicates the rate of decay of CDW amplitude.}
\end{figure}

To further characterize the CDW order, we have calculated the translational correlation. Calculating the correlation length of CDWs using topographic images can shed light on the interplay between CDWs and intercalant density. Non modulated boundaries of CDW modulated regions can be easily identified visually from STM topography images of areas with large defect concentration as in Fig.3\textbf{(a)}. Correlation length can be used to gauge the disruption of CDW order because of intercalation. Correlation length measurement can pick up the change in CDW order where it is not easily discernible from STM images visually. The presence of intercalants disprupts the CDW order hence results in a reduced translational correlation length. 

CDW peaks in the Fourier transform of topography images were selected and IFT was performed. IFT image displays CDW amplitudes as function of position (as in Fig.2\textbf{(d)} and 2\textbf{(g)}). Autocorrelation of this image provides information about the translational correlation length. The correlation length was obtained by calculating the change in amplitude of CDW from line sections in the auto correlation of IFT images, using following equation \cite{arguello2014visualizing},

\begin{equation}\label{eq1}
    A(x) = A_0 e^{-x/\xi}\cos(k_\text{cdw}x) + B_0
\end{equation}

Where $A(x)$ is CDW amplitude, $\xi$ is correlation length, $k_\text{cdw}$ wave vector of CDW in $x$ direction, $B_0$ is average random noise in autocorrelation obtained from IFT and $A_0 = 1 - B_0$. The finite correlation length is a consequence of the in-homogeneity and finite scanning area \cite{arguello2014visualizing}. Autocorrelation of STM topography image of area with more defects in Fig.2\textbf{(a)} is shown in Fig.4\textbf{(a)} and area with less number of defects in Fig.2\textbf{(e)}is shown in Fig.4\textbf{(b)}. Line section along the axis in autocorrelation image that is the amplitude of CDW as a function of $x$ was plotted. Correlation length was calculated by a fitting curve to the decay rate of amplitude in the line section.  In Fig.4\textbf{(b)} and \textbf{(d)} blue curve is line section along CDW direction indicates its amplitude and red curve fitted using (\ref{eq1}) shows rate of change of CDW amplitude. The correlation length obtained from fitting indicates that the correlation length is smaller for images with more intercalant sites than images with fewer intercalant sites. This implies the long-range coherence of CDWs has broken into small domains because of intercalation.  A sharp decline of CDW amplitudes means a smaller correlation length, which is a sign of the reduced long-range order. 

\begin{figure}[h]
\includegraphics[width=8cm]{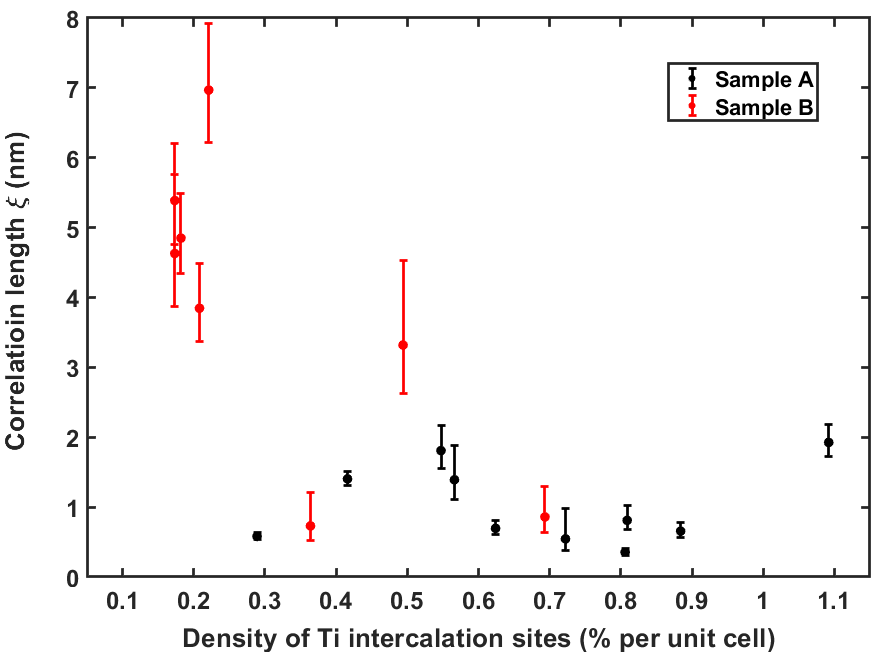}
\caption{Correlation length obtained by fitting the CDW amplitudes for topography images of different Ti intercalation concentration. Sample A contains more intercalation sites than sample B. (The average density of intercalation sites for sample A is $0.6\%$ and for sample B is $0.2\%$). Correlation length of CDWs in sample B seems to be higher than that of CDWs in sample A }
\end{figure}

The correlation length of CDW is a function of the concentration of Ti intercalation. A better picture of the exact dependence of correlation length on Ti intercalation density can be obtained by plotting the correlation length, as shown in  Fig.5. Atomically resolved images were taken, and the number of Ti intercalation sites were counted. The correlation length of CDWs was calculated from the autocorrelation of IFT images, as mentioned above. Ti intercalation concentration in Sample B is lesser than Sample A and shows a large correlation length.  The overall trend indicates that the correlation length of CDWs decreases as the Ti intercalation increases. The dependence of correlation length on electron doping nature of Ti intercalant strongly supports our hypothesis discussed in the earlier part of the excitonic origin of CDWs. However, there are certain features in the graph that require close examination. The correlation length tends to increase gradually at defect concentration of $0.3\%$, reaching maxima at $0.5\%$ then decrease again. This feature could be connected to the in-equivalent $Ti$ defect sites for each defect that are a consequence of CDW modulation \cite{novello2015scanning}. We believe our experimental result will motivate further detailed Density Functional theory (DFT) based calculations to explain this pattern in our correlation length.

\section{Conclusions}
We have explored defects at the atomic scale in two crystals of $1T$-TiSe$_2$ grown using different methods. Ti atoms get intercalated in the van der Waals gap. We find that the density of Ti intercalation affects the long-range order of CDWs. An increased concentration of Ti intercalation forces CDWs to form domains. There is no phase shift across the domain walls. The correlation length of CDWs is inversely proportional to the density of intercalated Ti atoms. Intercalation of Ti in $1T$-TiSe$_2$ suppresses CDWs and forms domain boundaries that can be explained by electron doping nature of Ti. Intercalated Ti atoms reduce the number of holes in the system. In the excitonic condensation model of CDW formation, the reduction of hole concentration depletes excitonic bound states. CDW suppression is a direct consequence of excitonic bound state depletion. Thus we demonstrate that our study of the dependence of correlation length on Ti intercalation supports the excitonic condensation model of formation of CDWs in $1T$-TiSe$_2$.

\begin{acknowledgments}
LH acknowledges the financial support from the Department of Science and Technology(DST), India [Grant No. SR/WOS-A/PM-33/2018 (G)] and IISER Pune. This work was financially supported by the Indian Institute of Science Education and Research (IISER), Pune. IM is grateful to CSIR, Government of India, for SRF and IISER Pune, for financial support in the form of a research fellowship, and UR is thankful to UGC, Government of India, for SRF.

\end{acknowledgments}

\bibliography{TiSe2_1}

\providecommand{\noopsort}[1]{}\providecommand{\singleletter}[1]{#1}%
\begin{thebibliography}{50}%
\makeatletter
\providecommand \@ifxundefined [1]{%
 \@ifx{#1\undefined}
}%
\providecommand \@ifnum [1]{%
 \ifnum #1\expandafter \@firstoftwo
 \else \expandafter \@secondoftwo
 \fi
}%
\providecommand \@ifx [1]{%
 \ifx #1\expandafter \@firstoftwo
 \else \expandafter \@secondoftwo
 \fi
}%
\providecommand \natexlab [1]{#1}%
\providecommand \enquote  [1]{``#1''}%
\providecommand \bibnamefont  [1]{#1}%
\providecommand \bibfnamefont [1]{#1}%
\providecommand \citenamefont [1]{#1}%
\providecommand \href@noop [0]{\@secondoftwo}%
\providecommand \href [0]{\begingroup \@sanitize@url \@href}%
\providecommand \@href[1]{\@@startlink{#1}\@@href}%
\providecommand \@@href[1]{\endgroup#1\@@endlink}%
\providecommand \@sanitize@url [0]{\catcode `\\12\catcode `\$12\catcode
  `\&12\catcode `\#12\catcode `\^12\catcode `\_12\catcode `\%12\relax}%
\providecommand \@@startlink[1]{}%
\providecommand \@@endlink[0]{}%
\providecommand \url  [0]{\begingroup\@sanitize@url \@url }%
\providecommand \@url [1]{\endgroup\@href {#1}{\urlprefix }}%
\providecommand \urlprefix  [0]{URL }%
\providecommand \Eprint [0]{\href }%
\providecommand \doibase [0]{https://doi.org/}%
\providecommand \selectlanguage [0]{\@gobble}%
\providecommand \bibinfo  [0]{\@secondoftwo}%
\providecommand \bibfield  [0]{\@secondoftwo}%
\providecommand \translation [1]{[#1]}%
\providecommand \BibitemOpen [0]{}%
\providecommand \bibitemStop [0]{}%
\providecommand \bibitemNoStop [0]{.\EOS\space}%
\providecommand \EOS [0]{\spacefactor3000\relax}%
\providecommand \BibitemShut  [1]{\csname bibitem#1\endcsname}%
\let\auto@bib@innerbib\@empty
%</preamble>
\bibitem [{\citenamefont {Tedstone}\ \emph {et~al.}(2016)\citenamefont
  {Tedstone}, \citenamefont {Lewis},\ and\ \citenamefont
  {O'Brien}}]{Tedstone2016}%
  \BibitemOpen
  \bibfield  {author} {\bibinfo {author} {\bibfnamefont {A.~A.}\ \bibnamefont
  {Tedstone}}, \bibinfo {author} {\bibfnamefont {D.~J.}\ \bibnamefont
  {Lewis}},\ and\ \bibinfo {author} {\bibfnamefont {P.}~\bibnamefont
  {O'Brien}},\ }\bibfield  {title} {\bibinfo {title} {{Synthesis, Properties,
  and Applications of Transition Metal-Doped Layered Transition Metal
  Dichalcogenides}},\ }\href {https://doi.org/10.1021/acs.chemmater.6b00430}
  {\bibfield  {journal} {\bibinfo  {journal} {Chemistry of Materials}\ }\textbf
  {\bibinfo {volume} {28}},\ \bibinfo {pages} {1965} (\bibinfo {year}
  {2016})}\BibitemShut {NoStop}%
\bibitem [{\citenamefont {Mueller}\ and\ \citenamefont
  {Malic}(2018)}]{mueller2018exciton}%
  \BibitemOpen
  \bibfield  {author} {\bibinfo {author} {\bibfnamefont {T.}~\bibnamefont
  {Mueller}}\ and\ \bibinfo {author} {\bibfnamefont {E.}~\bibnamefont
  {Malic}},\ }\bibfield  {title} {\bibinfo {title} {Exciton physics and device
  application of two-dimensional transition metal dichalcogenide
  semiconductors},\ }\href@noop {} {\bibfield  {journal} {\bibinfo  {journal}
  {NPJ 2D Materials and Applications}\ }\textbf {\bibinfo {volume} {2}},\
  \bibinfo {pages} {1} (\bibinfo {year} {2018})}\BibitemShut {NoStop}%
\bibitem [{\citenamefont {Pallecchi}\ \emph {et~al.}(2020)\citenamefont
  {Pallecchi}, \citenamefont {Manca}, \citenamefont {Patil}, \citenamefont
  {Pellegrino},\ and\ \citenamefont {Marre}}]{pallecchi2020review}%
  \BibitemOpen
  \bibfield  {author} {\bibinfo {author} {\bibfnamefont {I.}~\bibnamefont
  {Pallecchi}}, \bibinfo {author} {\bibfnamefont {N.}~\bibnamefont {Manca}},
  \bibinfo {author} {\bibfnamefont {B.}~\bibnamefont {Patil}}, \bibinfo
  {author} {\bibfnamefont {L.}~\bibnamefont {Pellegrino}},\ and\ \bibinfo
  {author} {\bibfnamefont {D.}~\bibnamefont {Marre}},\ }\bibfield  {title}
  {\bibinfo {title} {Review on thermoelectric properties of transition metal
  dichalcogenides},\ }\href@noop {} {\bibfield  {journal} {\bibinfo  {journal}
  {Nano Futures}\ } (\bibinfo {year} {2020})}\BibitemShut {NoStop}%
\bibitem [{\citenamefont {Huang}\ \emph {et~al.}(2020)\citenamefont {Huang},
  \citenamefont {Fan}, \citenamefont {Singh},\ and\ \citenamefont
  {Zheng}}]{huang2020recent}%
  \BibitemOpen
  \bibfield  {author} {\bibinfo {author} {\bibfnamefont {H.}~\bibnamefont
  {Huang}}, \bibinfo {author} {\bibfnamefont {X.}~\bibnamefont {Fan}}, \bibinfo
  {author} {\bibfnamefont {D.~J.}\ \bibnamefont {Singh}},\ and\ \bibinfo
  {author} {\bibfnamefont {W.~T.}\ \bibnamefont {Zheng}},\ }\bibfield  {title}
  {\bibinfo {title} {Recent progress of \textsc{TMD} nanomaterials: phase
  transitions and applications},\ }\href@noop {} {\bibfield  {journal}
  {\bibinfo  {journal} {Nanoscale}\ }\textbf {\bibinfo {volume} {12}},\
  \bibinfo {pages} {1247} (\bibinfo {year} {2020})}\BibitemShut {NoStop}%
\bibitem [{\citenamefont {Alexander V.~Kolobov}(2016)}]{book1}%
  \BibitemOpen
  \bibfield  {author} {\bibinfo {author} {\bibfnamefont {J.~T.}\ \bibnamefont
  {Alexander V.~Kolobov}},\ }\href@noop {} {\emph {\bibinfo {title} {Two
  Dimensional Transition metal dichalcogenides}}},\ \bibinfo {edition} {1st}\
  ed.,\ \bibinfo {series} {Series in Material Sciences}, Vol.\ \bibinfo
  {volume} {239}\ (\bibinfo  {publisher} {Springer},\ \bibinfo {year}
  {2016})\BibitemShut {NoStop}%
\bibitem [{\citenamefont {Isomaki}\ and\ \citenamefont
  {Von~Boehm}(1981)}]{isomaki1981gaps}%
  \BibitemOpen
  \bibfield  {author} {\bibinfo {author} {\bibfnamefont {H.}~\bibnamefont
  {Isomaki}}\ and\ \bibinfo {author} {\bibfnamefont {J.}~\bibnamefont
  {Von~Boehm}},\ }\bibfield  {title} {\bibinfo {title} {The gaps of the ideal
  \textsc{T}i\textsc{S}$_2$ and \textsc{T}i\textsc{S}e$_2$},\ }\href@noop {}
  {\bibfield  {journal} {\bibinfo  {journal} {Journal of Physics C: Solid State
  Physics}\ }\textbf {\bibinfo {volume} {14}},\ \bibinfo {pages} {L75}
  (\bibinfo {year} {1981})}\BibitemShut {NoStop}%
\bibitem [{\citenamefont {Di~Salvo}\ \emph {et~al.}(1976)\citenamefont
  {Di~Salvo}, \citenamefont {Moncton},\ and\ \citenamefont
  {Waszczak}}]{di1976electronic}%
  \BibitemOpen
  \bibfield  {author} {\bibinfo {author} {\bibfnamefont {F.~J.}\ \bibnamefont
  {Di~Salvo}}, \bibinfo {author} {\bibfnamefont {D.}~\bibnamefont {Moncton}},\
  and\ \bibinfo {author} {\bibfnamefont {J.}~\bibnamefont {Waszczak}},\
  }\bibfield  {title} {\bibinfo {title} {Electronic properties and superlattice
  formation in the semimetal \textsc{T}i\textsc{S}e$_2$},\ }\href@noop {}
  {\bibfield  {journal} {\bibinfo  {journal} {Physical Review B}\ }\textbf
  {\bibinfo {volume} {14}},\ \bibinfo {pages} {4321} (\bibinfo {year}
  {1976})}\BibitemShut {NoStop}%
\bibitem [{\citenamefont {Gr{\"u}ner}(1988)}]{gruner1988dynamics}%
  \BibitemOpen
  \bibfield  {author} {\bibinfo {author} {\bibfnamefont {G.}~\bibnamefont
  {Gr{\"u}ner}},\ }\bibfield  {title} {\bibinfo {title} {The dynamics of
  charge-density waves},\ }\href@noop {} {\bibfield  {journal} {\bibinfo
  {journal} {Reviews of modern physics}\ }\textbf {\bibinfo {volume} {60}},\
  \bibinfo {pages} {1129} (\bibinfo {year} {1988})}\BibitemShut {NoStop}%
\bibitem [{\citenamefont {Rossnagel}(2011)}]{rossnagel2011origin}%
  \BibitemOpen
  \bibfield  {author} {\bibinfo {author} {\bibfnamefont {K.}~\bibnamefont
  {Rossnagel}},\ }\bibfield  {title} {\bibinfo {title} {On the origin of
  charge-density waves in select layered transition-metal dichalcogenides},\
  }\href@noop {} {\bibfield  {journal} {\bibinfo  {journal} {Journal of
  Physics: Condensed Matter}\ }\textbf {\bibinfo {volume} {23}},\ \bibinfo
  {pages} {213001} (\bibinfo {year} {2011})}\BibitemShut {NoStop}%
\bibitem [{\citenamefont {Li}\ \emph {et~al.}(lack)\citenamefont {Li},
  \citenamefont {O’Farrell}, \citenamefont {Loh}, \citenamefont {Eda},
  \citenamefont {{\"O}zyilmaz},\ and\ \citenamefont
  {Neto}}]{li2016controlling}%
  \BibitemOpen
  \bibfield  {author} {\bibinfo {author} {\bibfnamefont {L.}~\bibnamefont
  {Li}}, \bibinfo {author} {\bibfnamefont {E.}~\bibnamefont {O’Farrell}},
  \bibinfo {author} {\bibfnamefont {K.}~\bibnamefont {Loh}}, \bibinfo {author}
  {\bibfnamefont {G.}~\bibnamefont {Eda}}, \bibinfo {author} {\bibfnamefont
  {B.}~\bibnamefont {{\"O}zyilmaz}},\ and\ \bibinfo {author} {\bibfnamefont
  {A.~C.}\ \bibnamefont {Neto}},\ }\bibfield  {title} {\bibinfo {title}
  {Controlling many-body states by the electric-field effect in a
  two-dimensional material},\ }\href@noop {} {\bibfield  {journal} {\bibinfo
  {journal} {Nature}\ }\textbf {\bibinfo {volume} {529}},\ \bibinfo {pages}
  {185} (\bibinfo {year} {2016\color{black}})}\BibitemShut {NoStop}%
\bibitem [{\citenamefont {Morosan}\ \emph {et~al.}(2006)\citenamefont
  {Morosan}, \citenamefont {Zandbergen}, \citenamefont {Dennis}, \citenamefont
  {Bos}, \citenamefont {Onose}, \citenamefont {Klimczuk}, \citenamefont
  {Ramirez}, \citenamefont {Ong},\ and\ \citenamefont
  {Cava}}]{morosan2006superconductivity}%
  \BibitemOpen
  \bibfield  {author} {\bibinfo {author} {\bibfnamefont {E.}~\bibnamefont
  {Morosan}}, \bibinfo {author} {\bibfnamefont {H.~W.}\ \bibnamefont
  {Zandbergen}}, \bibinfo {author} {\bibfnamefont {B.}~\bibnamefont {Dennis}},
  \bibinfo {author} {\bibfnamefont {J.}~\bibnamefont {Bos}}, \bibinfo {author}
  {\bibfnamefont {Y.}~\bibnamefont {Onose}}, \bibinfo {author} {\bibfnamefont
  {T.}~\bibnamefont {Klimczuk}}, \bibinfo {author} {\bibfnamefont
  {A.}~\bibnamefont {Ramirez}}, \bibinfo {author} {\bibfnamefont
  {N.}~\bibnamefont {Ong}},\ and\ \bibinfo {author} {\bibfnamefont {R.~J.}\
  \bibnamefont {Cava}},\ }\bibfield  {title} {\bibinfo {title}
  {Superconductivity in \textsc{C}u$_x$\textsc{T}i\textsc{S}e$_2$},\
  }\href@noop {} {\bibfield  {journal} {\bibinfo  {journal} {Nature Physics}\
  }\textbf {\bibinfo {volume} {2}},\ \bibinfo {pages} {544} (\bibinfo {year}
  {2006})}\BibitemShut {NoStop}%
\bibitem [{\citenamefont {Joe}\ \emph {et~al.}(2014)\citenamefont {Joe},
  \citenamefont {Chen}, \citenamefont {Ghaemi}, \citenamefont {Finkelstein},
  \citenamefont {de~La~Pe{\~n}a}, \citenamefont {Gan}, \citenamefont {Lee},
  \citenamefont {Yuan}, \citenamefont {Geck}, \citenamefont {MacDougall} \emph
  {et~al.}}]{joe2014emergence}%
  \BibitemOpen
  \bibfield  {author} {\bibinfo {author} {\bibfnamefont {Y.~I.}\ \bibnamefont
  {Joe}}, \bibinfo {author} {\bibfnamefont {X.}~\bibnamefont {Chen}}, \bibinfo
  {author} {\bibfnamefont {P.}~\bibnamefont {Ghaemi}}, \bibinfo {author}
  {\bibfnamefont {K.}~\bibnamefont {Finkelstein}}, \bibinfo {author}
  {\bibfnamefont {G.}~\bibnamefont {de~La~Pe{\~n}a}}, \bibinfo {author}
  {\bibfnamefont {Y.}~\bibnamefont {Gan}}, \bibinfo {author} {\bibfnamefont
  {J.}~\bibnamefont {Lee}}, \bibinfo {author} {\bibfnamefont {S.}~\bibnamefont
  {Yuan}}, \bibinfo {author} {\bibfnamefont {J.}~\bibnamefont {Geck}}, \bibinfo
  {author} {\bibfnamefont {G.}~\bibnamefont {MacDougall}}, \emph {et~al.},\
  }\bibfield  {title} {\bibinfo {title} {Emergence of charge density wave
  domain walls above the superconducting dome in
  \textit{1T}-\textsc{T}i\textsc{S}e$_2$},\ }\href@noop {} {\bibfield
  {journal} {\bibinfo  {journal} {Nature Physics}\ }\textbf {\bibinfo {volume}
  {10}},\ \bibinfo {pages} {421} (\bibinfo {year} {2014})}\BibitemShut
  {NoStop}%
\bibitem [{\citenamefont {Kogar}\ \emph
  {et~al.}(2017{\natexlab{a}})\citenamefont {Kogar}, \citenamefont
  {de~La~Pena}, \citenamefont {Lee}, \citenamefont {Fang}, \citenamefont {Sun},
  \citenamefont {Lioi}, \citenamefont {Karapetrov}, \citenamefont
  {Finkelstein}, \citenamefont {Ruff}, \citenamefont {Abbamonte} \emph
  {et~al.}}]{kogar2017observation}%
  \BibitemOpen
  \bibfield  {author} {\bibinfo {author} {\bibfnamefont {A.}~\bibnamefont
  {Kogar}}, \bibinfo {author} {\bibfnamefont {G.~A.}\ \bibnamefont
  {de~La~Pena}}, \bibinfo {author} {\bibfnamefont {S.}~\bibnamefont {Lee}},
  \bibinfo {author} {\bibfnamefont {Y.}~\bibnamefont {Fang}}, \bibinfo {author}
  {\bibfnamefont {S.-L.}\ \bibnamefont {Sun}}, \bibinfo {author} {\bibfnamefont
  {D.~B.}\ \bibnamefont {Lioi}}, \bibinfo {author} {\bibfnamefont
  {G.}~\bibnamefont {Karapetrov}}, \bibinfo {author} {\bibfnamefont {K.~D.}\
  \bibnamefont {Finkelstein}}, \bibinfo {author} {\bibfnamefont {J.~P.}\
  \bibnamefont {Ruff}}, \bibinfo {author} {\bibfnamefont {P.}~\bibnamefont
  {Abbamonte}}, \emph {et~al.},\ }\bibfield  {title} {\bibinfo {title}
  {Observation of a charge density wave incommensuration near the
  superconducting domain \textsc{C}u$_x$\textsc{T}i\textsc{S}e$_2$},\
  }\href@noop {} {\bibfield  {journal} {\bibinfo  {journal} {Physical review
  letters}\ }\textbf {\bibinfo {volume} {118}},\ \bibinfo {pages} {027002}
  (\bibinfo {year} {2017}{\natexlab{a}})}\BibitemShut {NoStop}%
\bibitem [{\citenamefont {J{\'e}rome}\ \emph {et~al.}(1967)\citenamefont
  {J{\'e}rome}, \citenamefont {Rice},\ and\ \citenamefont
  {Kohn}}]{jerome1967excitonic}%
  \BibitemOpen
  \bibfield  {author} {\bibinfo {author} {\bibfnamefont {D.}~\bibnamefont
  {J{\'e}rome}}, \bibinfo {author} {\bibfnamefont {T.}~\bibnamefont {Rice}},\
  and\ \bibinfo {author} {\bibfnamefont {W.}~\bibnamefont {Kohn}},\ }\bibfield
  {title} {\bibinfo {title} {Excitonic insulator},\ }\href@noop {} {\bibfield
  {journal} {\bibinfo  {journal} {Physical Review}\ }\textbf {\bibinfo {volume}
  {158}},\ \bibinfo {pages} {462} (\bibinfo {year} {1967})}\BibitemShut
  {NoStop}%
\bibitem [{\citenamefont {Hughes}(1977)}]{hughes1977structural}%
  \BibitemOpen
  \bibfield  {author} {\bibinfo {author} {\bibfnamefont {H.}~\bibnamefont
  {Hughes}},\ }\bibfield  {title} {\bibinfo {title} {Structural distortion in
  \textsc{T}i\textsc{S}e$_2$ and related materials-a possible
  \textsc{J}ahn-\textsc{T}eller effect?},\ }\href@noop {} {\bibfield  {journal}
  {\bibinfo  {journal} {Journal of Physics C: Solid State Physics}\ }\textbf
  {\bibinfo {volume} {10}},\ \bibinfo {pages} {L319} (\bibinfo {year}
  {1977})}\BibitemShut {NoStop}%
\bibitem [{\citenamefont {Kune{\v{s}}}(2015)}]{kunevs2015excitonic}%
  \BibitemOpen
  \bibfield  {author} {\bibinfo {author} {\bibfnamefont {J.}~\bibnamefont
  {Kune{\v{s}}}},\ }\bibfield  {title} {\bibinfo {title} {Excitonic
  condensation in systems of strongly correlated electrons},\ }\href@noop {}
  {\bibfield  {journal} {\bibinfo  {journal} {Journal of Physics: Condensed
  Matter}\ }\textbf {\bibinfo {volume} {27}},\ \bibinfo {pages} {333201}
  (\bibinfo {year} {2015})}\BibitemShut {NoStop}%
\bibitem [{\citenamefont {Seki}\ \emph {et~al.}(2014)\citenamefont {Seki},
  \citenamefont {Wakisaka}, \citenamefont {Kaneko}, \citenamefont {Toriyama},
  \citenamefont {Konishi}, \citenamefont {Sudayama}, \citenamefont {Saini},
  \citenamefont {Arita}, \citenamefont {Namatame}, \citenamefont {Taniguchi}
  \emph {et~al.}}]{seki2014excitonic}%
  \BibitemOpen
  \bibfield  {author} {\bibinfo {author} {\bibfnamefont {K.}~\bibnamefont
  {Seki}}, \bibinfo {author} {\bibfnamefont {Y.}~\bibnamefont {Wakisaka}},
  \bibinfo {author} {\bibfnamefont {T.}~\bibnamefont {Kaneko}}, \bibinfo
  {author} {\bibfnamefont {T.}~\bibnamefont {Toriyama}}, \bibinfo {author}
  {\bibfnamefont {T.}~\bibnamefont {Konishi}}, \bibinfo {author} {\bibfnamefont
  {T.}~\bibnamefont {Sudayama}}, \bibinfo {author} {\bibfnamefont
  {N.}~\bibnamefont {Saini}}, \bibinfo {author} {\bibfnamefont
  {M.}~\bibnamefont {Arita}}, \bibinfo {author} {\bibfnamefont
  {H.}~\bibnamefont {Namatame}}, \bibinfo {author} {\bibfnamefont
  {M.}~\bibnamefont {Taniguchi}}, \emph {et~al.},\ }\bibfield  {title}
  {\bibinfo {title} {Excitonic bose-einstein condensation in
  \textsc{T}a$_2$\textsc{N}i\textsc{S}e$_5$ above room temperature},\
  }\href@noop {} {\bibfield  {journal} {\bibinfo  {journal} {Physical Review
  B}\ }\textbf {\bibinfo {volume} {90}},\ \bibinfo {pages} {155116} (\bibinfo
  {year} {2014})}\BibitemShut {NoStop}%
\bibitem [{\citenamefont {Brinkman}\ and\ \citenamefont
  {Hilgenkamp}(2005)}]{brinkman2005electron}%
  \BibitemOpen
  \bibfield  {author} {\bibinfo {author} {\bibfnamefont {A.}~\bibnamefont
  {Brinkman}}\ and\ \bibinfo {author} {\bibfnamefont {H.}~\bibnamefont
  {Hilgenkamp}},\ }\bibfield  {title} {\bibinfo {title} {Electron-hole coupling
  in high-\textsc{T}$_c$ cuprate superconductors},\ }\href@noop {} {\bibfield
  {journal} {\bibinfo  {journal} {Physica C: Superconductivity}\ }\textbf
  {\bibinfo {volume} {422}},\ \bibinfo {pages} {71} (\bibinfo {year}
  {2005})}\BibitemShut {NoStop}%
\bibitem [{\citenamefont {Wang}\ \emph {et~al.}(2019)\citenamefont {Wang},
  \citenamefont {Rhodes}, \citenamefont {Watanabe}, \citenamefont {Taniguchi},
  \citenamefont {Hone}, \citenamefont {Shan},\ and\ \citenamefont
  {Mak}}]{wang2019evidence}%
  \BibitemOpen
  \bibfield  {author} {\bibinfo {author} {\bibfnamefont {Z.}~\bibnamefont
  {Wang}}, \bibinfo {author} {\bibfnamefont {D.~A.}\ \bibnamefont {Rhodes}},
  \bibinfo {author} {\bibfnamefont {K.}~\bibnamefont {Watanabe}}, \bibinfo
  {author} {\bibfnamefont {T.}~\bibnamefont {Taniguchi}}, \bibinfo {author}
  {\bibfnamefont {J.~C.}\ \bibnamefont {Hone}}, \bibinfo {author}
  {\bibfnamefont {J.}~\bibnamefont {Shan}},\ and\ \bibinfo {author}
  {\bibfnamefont {K.~F.}\ \bibnamefont {Mak}},\ }\bibfield  {title} {\bibinfo
  {title} {Evidence of high-temperature exciton condensation in two-dimensional
  atomic double layers},\ }\href@noop {} {\bibfield  {journal} {\bibinfo
  {journal} {Nature}\ }\textbf {\bibinfo {volume} {574}},\ \bibinfo {pages}
  {76} (\bibinfo {year} {2019})}\BibitemShut {NoStop}%
\bibitem [{\citenamefont {Shkvarin}\ \emph {et~al.}(2020)\citenamefont
  {Shkvarin}, \citenamefont {Merentsov}, \citenamefont {Tsud},\ and\
  \citenamefont {Titov}}]{shkvarin2020effect}%
  \BibitemOpen
  \bibfield  {author} {\bibinfo {author} {\bibfnamefont {A.~S.}\ \bibnamefont
  {Shkvarin}}, \bibinfo {author} {\bibfnamefont {A.~I.}\ \bibnamefont
  {Merentsov}}, \bibinfo {author} {\bibfnamefont {N.}~\bibnamefont {Tsud}},\
  and\ \bibinfo {author} {\bibfnamefont {A.~N.}\ \bibnamefont {Titov}},\
  }\bibfield  {title} {\bibinfo {title} {Effect of the titanium
  self-intercalation on the electronic structure of
  \textit{1T}-\textsc{T}i\textsc{S}e$_2$},\ }\href@noop {} {\bibfield
  {journal} {\bibinfo  {journal} {Inorganic Chemistry}\ } (\bibinfo {year}
  {2020})}\BibitemShut {NoStop}%
\bibitem [{\citenamefont {Krasavin}\ \emph {et~al.}(1998)\citenamefont
  {Krasavin}, \citenamefont {Titov},\ and\ \citenamefont
  {Antropov}}]{krasavin1998effect}%
  \BibitemOpen
  \bibfield  {author} {\bibinfo {author} {\bibfnamefont {L.}~\bibnamefont
  {Krasavin}}, \bibinfo {author} {\bibfnamefont {A.}~\bibnamefont {Titov}},\
  and\ \bibinfo {author} {\bibfnamefont {V.}~\bibnamefont {Antropov}},\
  }\bibfield  {title} {\bibinfo {title} {Effect of intercalation by silver on
  the charge-density-wave state in \textsc{T}i\textsc{S}e$_2$},\ }\href@noop {}
  {\bibfield  {journal} {\bibinfo  {journal} {Physics of the Solid State}\
  }\textbf {\bibinfo {volume} {40}},\ \bibinfo {pages} {1962} (\bibinfo {year}
  {1998})}\BibitemShut {NoStop}%
\bibitem [{\citenamefont {Watson}\ \emph {et~al.}(2019)\citenamefont {Watson},
  \citenamefont {Clark}, \citenamefont {Mazzola}, \citenamefont {Markovi{\'c}},
  \citenamefont {Sunko}, \citenamefont {Kim}, \citenamefont {Rossnagel},\ and\
  \citenamefont {King}}]{watson2019orbital}%
  \BibitemOpen
  \bibfield  {author} {\bibinfo {author} {\bibfnamefont {M.~D.}\ \bibnamefont
  {Watson}}, \bibinfo {author} {\bibfnamefont {O.~J.}\ \bibnamefont {Clark}},
  \bibinfo {author} {\bibfnamefont {F.}~\bibnamefont {Mazzola}}, \bibinfo
  {author} {\bibfnamefont {I.}~\bibnamefont {Markovi{\'c}}}, \bibinfo {author}
  {\bibfnamefont {V.}~\bibnamefont {Sunko}}, \bibinfo {author} {\bibfnamefont
  {T.~K.}\ \bibnamefont {Kim}}, \bibinfo {author} {\bibfnamefont
  {K.}~\bibnamefont {Rossnagel}},\ and\ \bibinfo {author} {\bibfnamefont
  {P.~D.}\ \bibnamefont {King}},\ }\bibfield  {title} {\bibinfo {title}
  {Orbital-and k$_z$ -selective hybridization of \textsc{S}e 4p and \textsc{T}i
  3d states in the charge density wave phase of \textsc{T}i\textsc{S}e$_2$},\
  }\href@noop {} {\bibfield  {journal} {\bibinfo  {journal} {Physical Review
  Letters}\ }\textbf {\bibinfo {volume} {122}},\ \bibinfo {pages} {076404}
  (\bibinfo {year} {2019})}\BibitemShut {NoStop}%
\bibitem [{\citenamefont {Chen}\ \emph
  {et~al.}(2015{\natexlab{a}})\citenamefont {Chen}, \citenamefont {Chan},
  \citenamefont {Fang}, \citenamefont {Zhang}, \citenamefont {Chou},
  \citenamefont {Mo}, \citenamefont {Hussain}, \citenamefont {Fedorov},\ and\
  \citenamefont {Chiang}}]{chen2015charge}%
  \BibitemOpen
  \bibfield  {author} {\bibinfo {author} {\bibfnamefont {P.}~\bibnamefont
  {Chen}}, \bibinfo {author} {\bibfnamefont {Y.-H.}\ \bibnamefont {Chan}},
  \bibinfo {author} {\bibfnamefont {X.-Y.}\ \bibnamefont {Fang}}, \bibinfo
  {author} {\bibfnamefont {Y.}~\bibnamefont {Zhang}}, \bibinfo {author}
  {\bibfnamefont {M.-Y.}\ \bibnamefont {Chou}}, \bibinfo {author}
  {\bibfnamefont {S.-K.}\ \bibnamefont {Mo}}, \bibinfo {author} {\bibfnamefont
  {Z.}~\bibnamefont {Hussain}}, \bibinfo {author} {\bibfnamefont {A.-V.}\
  \bibnamefont {Fedorov}},\ and\ \bibinfo {author} {\bibfnamefont {T.-C.}\
  \bibnamefont {Chiang}},\ }\bibfield  {title} {\bibinfo {title} {Charge
  density wave transition in single-layer titanium diselenide},\ }\href@noop {}
  {\bibfield  {journal} {\bibinfo  {journal} {Nature communications}\ }\textbf
  {\bibinfo {volume} {6}},\ \bibinfo {pages} {1} (\bibinfo {year}
  {2015}{\natexlab{a}})}\BibitemShut {NoStop}%
\bibitem [{\citenamefont {Rossnagel}\ \emph {et~al.}(2002)\citenamefont
  {Rossnagel}, \citenamefont {Kipp},\ and\ \citenamefont
  {Skibowski}}]{rossnagel2002charge}%
  \BibitemOpen
  \bibfield  {author} {\bibinfo {author} {\bibfnamefont {K.}~\bibnamefont
  {Rossnagel}}, \bibinfo {author} {\bibfnamefont {L.}~\bibnamefont {Kipp}},\
  and\ \bibinfo {author} {\bibfnamefont {M.}~\bibnamefont {Skibowski}},\
  }\bibfield  {title} {\bibinfo {title} {Charge-density-wave phase transition
  in \textit{1T}- \textsc{T}i\textsc{S}e$_2$: Excitonic insulator versus
  band-type jahn-teller mechanism},\ }\href@noop {} {\bibfield  {journal}
  {\bibinfo  {journal} {Physical Review B}\ }\textbf {\bibinfo {volume} {65}},\
  \bibinfo {pages} {235101} (\bibinfo {year} {2002})}\BibitemShut {NoStop}%
\bibitem [{\citenamefont {Cercellier}\ \emph {et~al.}(2007)\citenamefont
  {Cercellier}, \citenamefont {Monney}, \citenamefont {Clerc}, \citenamefont
  {Battaglia}, \citenamefont {Despont}, \citenamefont {Garnier}, \citenamefont
  {Beck}, \citenamefont {Aebi}, \citenamefont {Patthey}, \citenamefont {Berger}
  \emph {et~al.}}]{cercellier2007evidence}%
  \BibitemOpen
  \bibfield  {author} {\bibinfo {author} {\bibfnamefont {H.}~\bibnamefont
  {Cercellier}}, \bibinfo {author} {\bibfnamefont {C.}~\bibnamefont {Monney}},
  \bibinfo {author} {\bibfnamefont {F.}~\bibnamefont {Clerc}}, \bibinfo
  {author} {\bibfnamefont {C.}~\bibnamefont {Battaglia}}, \bibinfo {author}
  {\bibfnamefont {L.}~\bibnamefont {Despont}}, \bibinfo {author} {\bibfnamefont
  {M.}~\bibnamefont {Garnier}}, \bibinfo {author} {\bibfnamefont
  {H.}~\bibnamefont {Beck}}, \bibinfo {author} {\bibfnamefont {P.}~\bibnamefont
  {Aebi}}, \bibinfo {author} {\bibfnamefont {L.}~\bibnamefont {Patthey}},
  \bibinfo {author} {\bibfnamefont {H.}~\bibnamefont {Berger}}, \emph
  {et~al.},\ }\bibfield  {title} {\bibinfo {title} {Evidence for an excitonic
  insulator phase in \textit{1T}-\textsc{T}i\textsc{S}e$_2$},\ }\href@noop {}
  {\bibfield  {journal} {\bibinfo  {journal} {Physical Review Letters}\
  }\textbf {\bibinfo {volume} {99}},\ \bibinfo {pages} {146403} (\bibinfo
  {year} {2007})}\BibitemShut {NoStop}%
\bibitem [{\citenamefont {Hildebrand}\ \emph {et~al.}(2016)\citenamefont
  {Hildebrand}, \citenamefont {Jaouen}, \citenamefont {Didiot}, \citenamefont
  {Razzoli}, \citenamefont {Monney}, \citenamefont {Mottas}, \citenamefont
  {Ubaldini}, \citenamefont {Berger}, \citenamefont {Barreteau}, \citenamefont
  {Beck} \emph {et~al.}}]{hildebrand2016short}%
  \BibitemOpen
  \bibfield  {author} {\bibinfo {author} {\bibfnamefont {B.}~\bibnamefont
  {Hildebrand}}, \bibinfo {author} {\bibfnamefont {T.}~\bibnamefont {Jaouen}},
  \bibinfo {author} {\bibfnamefont {C.}~\bibnamefont {Didiot}}, \bibinfo
  {author} {\bibfnamefont {E.}~\bibnamefont {Razzoli}}, \bibinfo {author}
  {\bibfnamefont {G.}~\bibnamefont {Monney}}, \bibinfo {author} {\bibfnamefont
  {M.-L.}\ \bibnamefont {Mottas}}, \bibinfo {author} {\bibfnamefont
  {A.}~\bibnamefont {Ubaldini}}, \bibinfo {author} {\bibfnamefont
  {H.}~\bibnamefont {Berger}}, \bibinfo {author} {\bibfnamefont
  {C.}~\bibnamefont {Barreteau}}, \bibinfo {author} {\bibfnamefont
  {H.}~\bibnamefont {Beck}}, \emph {et~al.},\ }\bibfield  {title} {\bibinfo
  {title} {Short-range phase coherence and origin of the
  \textit{1T}-\textsc{T}i\textsc{S}e$_2$ charge density wave},\ }\href@noop {}
  {\bibfield  {journal} {\bibinfo  {journal} {Physical Review B}\ }\textbf
  {\bibinfo {volume} {93}},\ \bibinfo {pages} {125140} (\bibinfo {year}
  {2016})}\BibitemShut {NoStop}%
\bibitem [{\citenamefont {Monney}\ \emph {et~al.}(2009)\citenamefont {Monney},
  \citenamefont {Cercellier}, \citenamefont {Clerc}, \citenamefont {Battaglia},
  \citenamefont {Schwier}, \citenamefont {Didiot}, \citenamefont {Garnier},
  \citenamefont {Beck}, \citenamefont {Aebi}, \citenamefont {Berger} \emph
  {et~al.}}]{monney2009spontaneous}%
  \BibitemOpen
  \bibfield  {author} {\bibinfo {author} {\bibfnamefont {C.}~\bibnamefont
  {Monney}}, \bibinfo {author} {\bibfnamefont {H.}~\bibnamefont {Cercellier}},
  \bibinfo {author} {\bibfnamefont {F.}~\bibnamefont {Clerc}}, \bibinfo
  {author} {\bibfnamefont {C.}~\bibnamefont {Battaglia}}, \bibinfo {author}
  {\bibfnamefont {E.}~\bibnamefont {Schwier}}, \bibinfo {author} {\bibfnamefont
  {C.}~\bibnamefont {Didiot}}, \bibinfo {author} {\bibfnamefont {M.~G.}\
  \bibnamefont {Garnier}}, \bibinfo {author} {\bibfnamefont {H.}~\bibnamefont
  {Beck}}, \bibinfo {author} {\bibfnamefont {P.}~\bibnamefont {Aebi}}, \bibinfo
  {author} {\bibfnamefont {H.}~\bibnamefont {Berger}}, \emph {et~al.},\
  }\bibfield  {title} {\bibinfo {title} {Spontaneous exciton condensation in
  \textit{1T}-\textsc{T}i\textsc{S}e$_2$:\textsc{BCS}-like approach},\
  }\href@noop {} {\bibfield  {journal} {\bibinfo  {journal} {Physical Review
  B}\ }\textbf {\bibinfo {volume} {79}},\ \bibinfo {pages} {045116} (\bibinfo
  {year} {2009})}\BibitemShut {NoStop}%
\bibitem [{\citenamefont {Monney}\ \emph {et~al.}(2011)\citenamefont {Monney},
  \citenamefont {Battaglia}, \citenamefont {Cercellier}, \citenamefont {Aebi},\
  and\ \citenamefont {Beck}}]{monney2011exciton}%
  \BibitemOpen
  \bibfield  {author} {\bibinfo {author} {\bibfnamefont {C.}~\bibnamefont
  {Monney}}, \bibinfo {author} {\bibfnamefont {C.}~\bibnamefont {Battaglia}},
  \bibinfo {author} {\bibfnamefont {H.}~\bibnamefont {Cercellier}}, \bibinfo
  {author} {\bibfnamefont {P.}~\bibnamefont {Aebi}},\ and\ \bibinfo {author}
  {\bibfnamefont {H.}~\bibnamefont {Beck}},\ }\bibfield  {title} {\bibinfo
  {title} {Exciton condensation driving the periodic lattice distortion of
  \textit{1T}-\textsc{T}i\textsc{S}e$_2$},\ }\href@noop {} {\bibfield
  {journal} {\bibinfo  {journal} {Physical Review Letters}\ }\textbf {\bibinfo
  {volume} {106}},\ \bibinfo {pages} {106404} (\bibinfo {year}
  {2011})}\BibitemShut {NoStop}%
\bibitem [{\citenamefont {Cazzaniga}\ \emph {et~al.}(2012)\citenamefont
  {Cazzaniga}, \citenamefont {Cercellier}, \citenamefont {Holzmann},
  \citenamefont {Monney}, \citenamefont {Aebi}, \citenamefont {Onida},\ and\
  \citenamefont {Olevano}}]{cazzaniga2012ab}%
  \BibitemOpen
  \bibfield  {author} {\bibinfo {author} {\bibfnamefont {M.}~\bibnamefont
  {Cazzaniga}}, \bibinfo {author} {\bibfnamefont {H.}~\bibnamefont
  {Cercellier}}, \bibinfo {author} {\bibfnamefont {M.}~\bibnamefont
  {Holzmann}}, \bibinfo {author} {\bibfnamefont {C.}~\bibnamefont {Monney}},
  \bibinfo {author} {\bibfnamefont {P.}~\bibnamefont {Aebi}}, \bibinfo {author}
  {\bibfnamefont {G.}~\bibnamefont {Onida}},\ and\ \bibinfo {author}
  {\bibfnamefont {V.}~\bibnamefont {Olevano}},\ }\bibfield  {title} {\bibinfo
  {title} {\textit{Ab initio} many-body effects in
  \textit{1T}-\textsc{T}i\textsc{S}e$_2$: A possible excitonic insulator
  scenario from gw band-shape renormalization},\ }\href@noop {} {\bibfield
  {journal} {\bibinfo  {journal} {Physical Review B}\ }\textbf {\bibinfo
  {volume} {85}},\ \bibinfo {pages} {195111} (\bibinfo {year}
  {2012})}\BibitemShut {NoStop}%
\bibitem [{\citenamefont {Kogar}\ \emph
  {et~al.}(2017{\natexlab{b}})\citenamefont {Kogar}, \citenamefont {Rak},
  \citenamefont {Vig}, \citenamefont {Husain}, \citenamefont {Flicker},
  \citenamefont {Joe}, \citenamefont {Venema}, \citenamefont {MacDougall},
  \citenamefont {Chiang}, \citenamefont {Fradkin} \emph
  {et~al.}}]{kogar2017signatures}%
  \BibitemOpen
  \bibfield  {author} {\bibinfo {author} {\bibfnamefont {A.}~\bibnamefont
  {Kogar}}, \bibinfo {author} {\bibfnamefont {M.~S.}\ \bibnamefont {Rak}},
  \bibinfo {author} {\bibfnamefont {S.}~\bibnamefont {Vig}}, \bibinfo {author}
  {\bibfnamefont {A.~A.}\ \bibnamefont {Husain}}, \bibinfo {author}
  {\bibfnamefont {F.}~\bibnamefont {Flicker}}, \bibinfo {author} {\bibfnamefont
  {Y.~I.}\ \bibnamefont {Joe}}, \bibinfo {author} {\bibfnamefont
  {L.}~\bibnamefont {Venema}}, \bibinfo {author} {\bibfnamefont {G.~J.}\
  \bibnamefont {MacDougall}}, \bibinfo {author} {\bibfnamefont {T.~C.}\
  \bibnamefont {Chiang}}, \bibinfo {author} {\bibfnamefont {E.}~\bibnamefont
  {Fradkin}}, \emph {et~al.},\ }\bibfield  {title} {\bibinfo {title}
  {Signatures of exciton condensation in a transition metal dichalcogenide},\
  }\href@noop {} {\bibfield  {journal} {\bibinfo  {journal} {Science}\ }\textbf
  {\bibinfo {volume} {358}},\ \bibinfo {pages} {1314} (\bibinfo {year}
  {2017}{\natexlab{b}})}\BibitemShut {NoStop}%
\bibitem [{\citenamefont {Snoke}(2002)}]{snoke2002spontaneous}%
  \BibitemOpen
  \bibfield  {author} {\bibinfo {author} {\bibfnamefont {D.}~\bibnamefont
  {Snoke}},\ }\bibfield  {title} {\bibinfo {title} {Spontaneous bose coherence
  of excitons and polaritons},\ }\href@noop {} {\bibfield  {journal} {\bibinfo
  {journal} {Science}\ }\textbf {\bibinfo {volume} {298}},\ \bibinfo {pages}
  {1368} (\bibinfo {year} {2002})}\BibitemShut {NoStop}%
\bibitem [{\citenamefont {Kohn}\ and\ \citenamefont
  {Sherrington}(1970)}]{kohn1970two}%
  \BibitemOpen
  \bibfield  {author} {\bibinfo {author} {\bibfnamefont {W.}~\bibnamefont
  {Kohn}}\ and\ \bibinfo {author} {\bibfnamefont {D.}~\bibnamefont
  {Sherrington}},\ }\bibfield  {title} {\bibinfo {title} {Two kinds of bosons
  and bose condensates},\ }\href@noop {} {\bibfield  {journal} {\bibinfo
  {journal} {Reviews of Modern Physics}\ }\textbf {\bibinfo {volume} {42}},\
  \bibinfo {pages} {1} (\bibinfo {year} {1970})}\BibitemShut {NoStop}%
\bibitem [{\citenamefont {Ishioka}\ \emph {et~al.}(2010)\citenamefont
  {Ishioka}, \citenamefont {Liu}, \citenamefont {Shimatake}, \citenamefont
  {Kurosawa}, \citenamefont {Ichimura}, \citenamefont {Toda}, \citenamefont
  {Oda},\ and\ \citenamefont {Tanda}}]{ishioka2010chiral}%
  \BibitemOpen
  \bibfield  {author} {\bibinfo {author} {\bibfnamefont {J.}~\bibnamefont
  {Ishioka}}, \bibinfo {author} {\bibfnamefont {Y.}~\bibnamefont {Liu}},
  \bibinfo {author} {\bibfnamefont {K.}~\bibnamefont {Shimatake}}, \bibinfo
  {author} {\bibfnamefont {T.}~\bibnamefont {Kurosawa}}, \bibinfo {author}
  {\bibfnamefont {K.}~\bibnamefont {Ichimura}}, \bibinfo {author}
  {\bibfnamefont {Y.}~\bibnamefont {Toda}}, \bibinfo {author} {\bibfnamefont
  {M.}~\bibnamefont {Oda}},\ and\ \bibinfo {author} {\bibfnamefont
  {S.}~\bibnamefont {Tanda}},\ }\bibfield  {title} {\bibinfo {title} {Chiral
  charge-density waves},\ }\href@noop {} {\bibfield  {journal} {\bibinfo
  {journal} {Physical Review Letters}\ }\textbf {\bibinfo {volume} {105}},\
  \bibinfo {pages} {176401} (\bibinfo {year} {2010})}\BibitemShut {NoStop}%
\bibitem [{\citenamefont {Ishioka}\ \emph {et~al.}(2011)\citenamefont
  {Ishioka}, \citenamefont {Fujii}, \citenamefont {Katono}, \citenamefont
  {Ichimura}, \citenamefont {Kurosawa}, \citenamefont {Oda},\ and\
  \citenamefont {Tanda}}]{ishioka2011charge}%
  \BibitemOpen
  \bibfield  {author} {\bibinfo {author} {\bibfnamefont {J.}~\bibnamefont
  {Ishioka}}, \bibinfo {author} {\bibfnamefont {T.}~\bibnamefont {Fujii}},
  \bibinfo {author} {\bibfnamefont {K.}~\bibnamefont {Katono}}, \bibinfo
  {author} {\bibfnamefont {K.}~\bibnamefont {Ichimura}}, \bibinfo {author}
  {\bibfnamefont {T.}~\bibnamefont {Kurosawa}}, \bibinfo {author}
  {\bibfnamefont {M.}~\bibnamefont {Oda}},\ and\ \bibinfo {author}
  {\bibfnamefont {S.}~\bibnamefont {Tanda}},\ }\bibfield  {title} {\bibinfo
  {title} {Charge-parity symmetry observed through friedel oscillations in
  chiral charge-density waves},\ }\href@noop {} {\bibfield  {journal} {\bibinfo
   {journal} {Physical Review B}\ }\textbf {\bibinfo {volume} {84}},\ \bibinfo
  {pages} {245125} (\bibinfo {year} {2011})}\BibitemShut {NoStop}%
\bibitem [{\citenamefont {Iavarone}\ \emph {et~al.}(2012)\citenamefont
  {Iavarone}, \citenamefont {Di~Capua}, \citenamefont {Zhang}, \citenamefont
  {Golalikhani}, \citenamefont {Moore},\ and\ \citenamefont
  {Karapetrov}}]{iavarone2012evolution}%
  \BibitemOpen
  \bibfield  {author} {\bibinfo {author} {\bibfnamefont {M.}~\bibnamefont
  {Iavarone}}, \bibinfo {author} {\bibfnamefont {R.}~\bibnamefont {Di~Capua}},
  \bibinfo {author} {\bibfnamefont {X.}~\bibnamefont {Zhang}}, \bibinfo
  {author} {\bibfnamefont {M.}~\bibnamefont {Golalikhani}}, \bibinfo {author}
  {\bibfnamefont {S.}~\bibnamefont {Moore}},\ and\ \bibinfo {author}
  {\bibfnamefont {G.}~\bibnamefont {Karapetrov}},\ }\bibfield  {title}
  {\bibinfo {title} {Evolution of the charge density wave state in
  \textsc{C}u$_x$\textsc{T}i\textsc{S}e$_2$},\ }\href@noop {} {\bibfield
  {journal} {\bibinfo  {journal} {Physical Review B}\ }\textbf {\bibinfo
  {volume} {85}},\ \bibinfo {pages} {155103} (\bibinfo {year}
  {2012})}\BibitemShut {NoStop}%
\bibitem [{\citenamefont {Kusmartseva}\ \emph {et~al.}(2009)\citenamefont
  {Kusmartseva}, \citenamefont {Sipos}, \citenamefont {Berger}, \citenamefont
  {Forro},\ and\ \citenamefont {Tuti{\v{s}}}}]{kusmartseva2009pressure}%
  \BibitemOpen
  \bibfield  {author} {\bibinfo {author} {\bibfnamefont {A.~F.}\ \bibnamefont
  {Kusmartseva}}, \bibinfo {author} {\bibfnamefont {B.}~\bibnamefont {Sipos}},
  \bibinfo {author} {\bibfnamefont {H.}~\bibnamefont {Berger}}, \bibinfo
  {author} {\bibfnamefont {L.}~\bibnamefont {Forro}},\ and\ \bibinfo {author}
  {\bibfnamefont {E.}~\bibnamefont {Tuti{\v{s}}}},\ }\bibfield  {title}
  {\bibinfo {title} {Pressure induced superconductivity in pristine
  \textit{1T}-\textsc{T}i\textsc{S}e$_2$},\ }\href@noop {} {\bibfield
  {journal} {\bibinfo  {journal} {Physical review letters}\ }\textbf {\bibinfo
  {volume} {103}},\ \bibinfo {pages} {236401} (\bibinfo {year}
  {2009})}\BibitemShut {NoStop}%
\bibitem [{\citenamefont {Morosan}\ \emph {et~al.}(2010)\citenamefont
  {Morosan}, \citenamefont {Wagner}, \citenamefont {Zhao}, \citenamefont {Hor},
  \citenamefont {Williams}, \citenamefont {Tao}, \citenamefont {Zhu},\ and\
  \citenamefont {Cava}}]{morosan2010multiple}%
  \BibitemOpen
  \bibfield  {author} {\bibinfo {author} {\bibfnamefont {E.}~\bibnamefont
  {Morosan}}, \bibinfo {author} {\bibfnamefont {K.~E.}\ \bibnamefont {Wagner}},
  \bibinfo {author} {\bibfnamefont {L.~L.}\ \bibnamefont {Zhao}}, \bibinfo
  {author} {\bibfnamefont {Y.}~\bibnamefont {Hor}}, \bibinfo {author}
  {\bibfnamefont {A.~J.}\ \bibnamefont {Williams}}, \bibinfo {author}
  {\bibfnamefont {J.}~\bibnamefont {Tao}}, \bibinfo {author} {\bibfnamefont
  {Y.}~\bibnamefont {Zhu}},\ and\ \bibinfo {author} {\bibfnamefont {R.~J.}\
  \bibnamefont {Cava}},\ }\bibfield  {title} {\bibinfo {title} {Multiple
  electronic transitions and superconductivity in
  \textsc{p}d$_x$\textsc{T}i\textsc{S}e$_2$},\ }\href@noop {} {\bibfield
  {journal} {\bibinfo  {journal} {Physical Review B}\ }\textbf {\bibinfo
  {volume} {81}},\ \bibinfo {pages} {094524} (\bibinfo {year}
  {2010})}\BibitemShut {NoStop}%
\bibitem [{\citenamefont {Chen}\ \emph
  {et~al.}(2015{\natexlab{b}})\citenamefont {Chen}, \citenamefont {Wang},
  \citenamefont {Carr}, \citenamefont {Vogel}, \citenamefont {Gourdon},
  \citenamefont {Dai},\ and\ \citenamefont {Morosan}}]{chen2015chemical}%
  \BibitemOpen
  \bibfield  {author} {\bibinfo {author} {\bibfnamefont {J.~S.}\ \bibnamefont
  {Chen}}, \bibinfo {author} {\bibfnamefont {J.~K.}\ \bibnamefont {Wang}},
  \bibinfo {author} {\bibfnamefont {S.~V.}\ \bibnamefont {Carr}}, \bibinfo
  {author} {\bibfnamefont {S.~C.}\ \bibnamefont {Vogel}}, \bibinfo {author}
  {\bibfnamefont {O.}~\bibnamefont {Gourdon}}, \bibinfo {author} {\bibfnamefont
  {P.}~\bibnamefont {Dai}},\ and\ \bibinfo {author} {\bibfnamefont
  {E.}~\bibnamefont {Morosan}},\ }\bibfield  {title} {\bibinfo {title}
  {Chemical tuning of electrical transport in
  \textsc{T}i$_{1-x}$\textsc{P}t$_x$\textsc{S}e$_{2-y}$},\ }\href@noop {}
  {\bibfield  {journal} {\bibinfo  {journal} {Physical Review B}\ }\textbf
  {\bibinfo {volume} {91}},\ \bibinfo {pages} {045125} (\bibinfo {year}
  {2015}{\natexlab{b}})}\BibitemShut {NoStop}%
\bibitem [{2Ds(2018)}]{2Dsemi}%
  \BibitemOpen
  \href@noop {} {\bibinfo {title} {{2D Semiconductors, USA}}},\ \bibinfo
  {howpublished} {\url{https://www.2dsemiconductors.com/}} (\bibinfo {year}
  {2018})\BibitemShut {NoStop}%
\bibitem [{\citenamefont {Slough}\ \emph {et~al.}(1988)\citenamefont {Slough},
  \citenamefont {Giambattista}, \citenamefont {Johnson}, \citenamefont
  {McNairy}, \citenamefont {Wang},\ and\ \citenamefont
  {Coleman}}]{slough1988scanning}%
  \BibitemOpen
  \bibfield  {author} {\bibinfo {author} {\bibfnamefont {C.}~\bibnamefont
  {Slough}}, \bibinfo {author} {\bibfnamefont {B.}~\bibnamefont
  {Giambattista}}, \bibinfo {author} {\bibfnamefont {A.}~\bibnamefont
  {Johnson}}, \bibinfo {author} {\bibfnamefont {W.}~\bibnamefont {McNairy}},
  \bibinfo {author} {\bibfnamefont {C.}~\bibnamefont {Wang}},\ and\ \bibinfo
  {author} {\bibfnamefont {R.}~\bibnamefont {Coleman}},\ }\bibfield  {title}
  {\bibinfo {title} {Scanning tunneling microscopy of
  \textit{1T}-\textsc{T}i\textsc{S}e$_2$ and
  \textit{1T}-\textsc{T}i\textsc{S}$_2$ at 77 and 4.2 k},\ }\href@noop {}
  {\bibfield  {journal} {\bibinfo  {journal} {Physical Review B}\ }\textbf
  {\bibinfo {volume} {37}},\ \bibinfo {pages} {6571} (\bibinfo {year}
  {1988})}\BibitemShut {NoStop}%
\bibitem [{\citenamefont {Hildebrand}\ \emph {et~al.}(2014)\citenamefont
  {Hildebrand}, \citenamefont {Didiot}, \citenamefont {Novello}, \citenamefont
  {Monney}, \citenamefont {Scarfato}, \citenamefont {Ubaldini}, \citenamefont
  {Berger}, \citenamefont {Bowler}, \citenamefont {Renner},\ and\ \citenamefont
  {Aebi}}]{hildebrand2014doping}%
  \BibitemOpen
  \bibfield  {author} {\bibinfo {author} {\bibfnamefont {B.}~\bibnamefont
  {Hildebrand}}, \bibinfo {author} {\bibfnamefont {C.}~\bibnamefont {Didiot}},
  \bibinfo {author} {\bibfnamefont {A.~M.}\ \bibnamefont {Novello}}, \bibinfo
  {author} {\bibfnamefont {G.}~\bibnamefont {Monney}}, \bibinfo {author}
  {\bibfnamefont {A.}~\bibnamefont {Scarfato}}, \bibinfo {author}
  {\bibfnamefont {A.}~\bibnamefont {Ubaldini}}, \bibinfo {author}
  {\bibfnamefont {H.}~\bibnamefont {Berger}}, \bibinfo {author} {\bibfnamefont
  {D.}~\bibnamefont {Bowler}}, \bibinfo {author} {\bibfnamefont
  {C.}~\bibnamefont {Renner}},\ and\ \bibinfo {author} {\bibfnamefont
  {P.}~\bibnamefont {Aebi}},\ }\bibfield  {title} {\bibinfo {title} {Doping
  nature of native defects in \textit{1T}-\textsc{T}i\textsc{S}e$_2$},\
  }\href@noop {} {\bibfield  {journal} {\bibinfo  {journal} {Physical review
  letters}\ }\textbf {\bibinfo {volume} {112}},\ \bibinfo {pages} {197001}
  (\bibinfo {year} {2014})}\BibitemShut {NoStop}%
\bibitem [{\citenamefont {Fan}\ \emph {et~al.}(2017)\citenamefont {Fan},
  \citenamefont {Zou}, \citenamefont {Du}, \citenamefont {Gan}, \citenamefont
  {Xu}, \citenamefont {Lv}, \citenamefont {He}, \citenamefont {Yang},
  \citenamefont {Kang},\ and\ \citenamefont {Li}}]{fan2017theoretical}%
  \BibitemOpen
  \bibfield  {author} {\bibinfo {author} {\bibfnamefont {S.}~\bibnamefont
  {Fan}}, \bibinfo {author} {\bibfnamefont {X.}~\bibnamefont {Zou}}, \bibinfo
  {author} {\bibfnamefont {H.}~\bibnamefont {Du}}, \bibinfo {author}
  {\bibfnamefont {L.}~\bibnamefont {Gan}}, \bibinfo {author} {\bibfnamefont
  {C.}~\bibnamefont {Xu}}, \bibinfo {author} {\bibfnamefont {W.}~\bibnamefont
  {Lv}}, \bibinfo {author} {\bibfnamefont {Y.-B.}\ \bibnamefont {He}}, \bibinfo
  {author} {\bibfnamefont {Q.-H.}\ \bibnamefont {Yang}}, \bibinfo {author}
  {\bibfnamefont {F.}~\bibnamefont {Kang}},\ and\ \bibinfo {author}
  {\bibfnamefont {J.}~\bibnamefont {Li}},\ }\bibfield  {title} {\bibinfo
  {title} {Theoretical investigation of the intercalation chemistry of
  lithium/sodium ions in transition metal dichalcogenides},\ }\href@noop {}
  {\bibfield  {journal} {\bibinfo  {journal} {The Journal of Physical Chemistry
  C}\ }\textbf {\bibinfo {volume} {121}},\ \bibinfo {pages} {13599} (\bibinfo
  {year} {2017})}\BibitemShut {NoStop}%
\bibitem [{\citenamefont {May}\ \emph {et~al.}(2011)\citenamefont {May},
  \citenamefont {Brabetz}, \citenamefont {Janowitz},\ and\ \citenamefont
  {Manzke}}]{may2011influence}%
  \BibitemOpen
  \bibfield  {author} {\bibinfo {author} {\bibfnamefont {M.~M.}\ \bibnamefont
  {May}}, \bibinfo {author} {\bibfnamefont {C.}~\bibnamefont {Brabetz}},
  \bibinfo {author} {\bibfnamefont {C.}~\bibnamefont {Janowitz}},\ and\
  \bibinfo {author} {\bibfnamefont {R.}~\bibnamefont {Manzke}},\ }\bibfield
  {title} {\bibinfo {title} {The influence of different growth conditions on
  the charge density wave transition of \textsc{T}i\textsc{S}e$_2$},\
  }\href@noop {} {\bibfield  {journal} {\bibinfo  {journal} {Journal of
  Electron Spectroscopy and Related Phenomena}\ }\textbf {\bibinfo {volume}
  {184}},\ \bibinfo {pages} {180} (\bibinfo {year} {2011})}\BibitemShut
  {NoStop}%
\bibitem [{\citenamefont {Sun}\ \emph {et~al.}(2017)\citenamefont {Sun},
  \citenamefont {Chen}, \citenamefont {Zhang}, \citenamefont {Sohrt},
  \citenamefont {Zhao}, \citenamefont {Xu}, \citenamefont {Wang}, \citenamefont
  {Wang}, \citenamefont {Rossnagel}, \citenamefont {Gu} \emph
  {et~al.}}]{sun2017suppression}%
  \BibitemOpen
  \bibfield  {author} {\bibinfo {author} {\bibfnamefont {L.}~\bibnamefont
  {Sun}}, \bibinfo {author} {\bibfnamefont {C.}~\bibnamefont {Chen}}, \bibinfo
  {author} {\bibfnamefont {Q.}~\bibnamefont {Zhang}}, \bibinfo {author}
  {\bibfnamefont {C.}~\bibnamefont {Sohrt}}, \bibinfo {author} {\bibfnamefont
  {T.}~\bibnamefont {Zhao}}, \bibinfo {author} {\bibfnamefont {G.}~\bibnamefont
  {Xu}}, \bibinfo {author} {\bibfnamefont {J.}~\bibnamefont {Wang}}, \bibinfo
  {author} {\bibfnamefont {D.}~\bibnamefont {Wang}}, \bibinfo {author}
  {\bibfnamefont {K.}~\bibnamefont {Rossnagel}}, \bibinfo {author}
  {\bibfnamefont {L.}~\bibnamefont {Gu}}, \emph {et~al.},\ }\bibfield  {title}
  {\bibinfo {title} {Suppression of the charge density wave state in
  two-dimensional \textit{1T}-\textsc{T}i\textsc{S}e$_2$ by atmospheric
  oxidation},\ }\href@noop {} {\bibfield  {journal} {\bibinfo  {journal}
  {Angewandte Chemie International Edition}\ }\textbf {\bibinfo {volume}
  {56}},\ \bibinfo {pages} {8981} (\bibinfo {year} {2017})}\BibitemShut
  {NoStop}%
\bibitem [{\citenamefont {Jaouen}\ \emph {et~al.}(2019)\citenamefont {Jaouen},
  \citenamefont {Hildebrand}, \citenamefont {Mottas}, \citenamefont
  {Di~Giovannantonio}, \citenamefont {Ruffieux}, \citenamefont {Rumo},
  \citenamefont {Nicholson}, \citenamefont {Razzoli}, \citenamefont
  {Barreteau}, \citenamefont {Ubaldini} \emph {et~al.}}]{jaouen2019phase}%
  \BibitemOpen
  \bibfield  {author} {\bibinfo {author} {\bibfnamefont {T.}~\bibnamefont
  {Jaouen}}, \bibinfo {author} {\bibfnamefont {B.}~\bibnamefont {Hildebrand}},
  \bibinfo {author} {\bibfnamefont {M.-L.}\ \bibnamefont {Mottas}}, \bibinfo
  {author} {\bibfnamefont {M.}~\bibnamefont {Di~Giovannantonio}}, \bibinfo
  {author} {\bibfnamefont {P.}~\bibnamefont {Ruffieux}}, \bibinfo {author}
  {\bibfnamefont {M.}~\bibnamefont {Rumo}}, \bibinfo {author} {\bibfnamefont
  {C.~W.}\ \bibnamefont {Nicholson}}, \bibinfo {author} {\bibfnamefont
  {E.}~\bibnamefont {Razzoli}}, \bibinfo {author} {\bibfnamefont
  {C.}~\bibnamefont {Barreteau}}, \bibinfo {author} {\bibfnamefont
  {A.}~\bibnamefont {Ubaldini}}, \emph {et~al.},\ }\bibfield  {title} {\bibinfo
  {title} {Phase separation in the vicinity of \textsc{F}ermi surface hot
  spots},\ }\href@noop {} {\bibfield  {journal} {\bibinfo  {journal} {Physical
  Review B}\ }\textbf {\bibinfo {volume} {100}},\ \bibinfo {pages} {075152}
  (\bibinfo {year} {2019})}\BibitemShut {NoStop}%
\bibitem [{\citenamefont {Yan}\ \emph {et~al.}(2017)\citenamefont {Yan},
  \citenamefont {Iaia}, \citenamefont {Morosan}, \citenamefont {Fradkin},
  \citenamefont {Abbamonte},\ and\ \citenamefont
  {Madhavan}}]{yan2017influence}%
  \BibitemOpen
  \bibfield  {author} {\bibinfo {author} {\bibfnamefont {S.}~\bibnamefont
  {Yan}}, \bibinfo {author} {\bibfnamefont {D.}~\bibnamefont {Iaia}}, \bibinfo
  {author} {\bibfnamefont {E.}~\bibnamefont {Morosan}}, \bibinfo {author}
  {\bibfnamefont {E.}~\bibnamefont {Fradkin}}, \bibinfo {author} {\bibfnamefont
  {P.}~\bibnamefont {Abbamonte}},\ and\ \bibinfo {author} {\bibfnamefont
  {V.}~\bibnamefont {Madhavan}},\ }\bibfield  {title} {\bibinfo {title}
  {Influence of domain walls in the incommensurate charge density wave state of
  \textsc{C}u intercalated \textit{1T}-\textsc{T}i\textsc{S}e$_2$},\
  }\href@noop {} {\bibfield  {journal} {\bibinfo  {journal} {Physical Review
  Letters}\ }\textbf {\bibinfo {volume} {118}},\ \bibinfo {pages} {106405}
  (\bibinfo {year} {2017})}\BibitemShut {NoStop}%
\bibitem [{\citenamefont {Jishi}\ and\ \citenamefont
  {Alyahyaei}(2008)}]{jishi2008electronic}%
  \BibitemOpen
  \bibfield  {author} {\bibinfo {author} {\bibfnamefont {R.}~\bibnamefont
  {Jishi}}\ and\ \bibinfo {author} {\bibfnamefont {H.}~\bibnamefont
  {Alyahyaei}},\ }\bibfield  {title} {\bibinfo {title} {Electronic structure of
  superconducting copper intercalated transition metal dichalcogenides:
  First-principles calculations},\ }\href@noop {} {\bibfield  {journal}
  {\bibinfo  {journal} {Physical Review B}\ }\textbf {\bibinfo {volume} {78}},\
  \bibinfo {pages} {144516} (\bibinfo {year} {2008})}\BibitemShut {NoStop}%
\bibitem [{\citenamefont {Novello}\ \emph {et~al.}(2015)\citenamefont
  {Novello}, \citenamefont {Hildebrand}, \citenamefont {Scarfato},
  \citenamefont {Didiot}, \citenamefont {Monney}, \citenamefont {Ubaldini},
  \citenamefont {Berger}, \citenamefont {Bowler}, \citenamefont {Aebi},\ and\
  \citenamefont {Renner}}]{novello2015scanning}%
  \BibitemOpen
  \bibfield  {author} {\bibinfo {author} {\bibfnamefont {A.~M.}\ \bibnamefont
  {Novello}}, \bibinfo {author} {\bibfnamefont {B.}~\bibnamefont {Hildebrand}},
  \bibinfo {author} {\bibfnamefont {A.}~\bibnamefont {Scarfato}}, \bibinfo
  {author} {\bibfnamefont {C.}~\bibnamefont {Didiot}}, \bibinfo {author}
  {\bibfnamefont {G.}~\bibnamefont {Monney}}, \bibinfo {author} {\bibfnamefont
  {A.}~\bibnamefont {Ubaldini}}, \bibinfo {author} {\bibfnamefont
  {H.}~\bibnamefont {Berger}}, \bibinfo {author} {\bibfnamefont
  {D.}~\bibnamefont {Bowler}}, \bibinfo {author} {\bibfnamefont
  {P.}~\bibnamefont {Aebi}},\ and\ \bibinfo {author} {\bibfnamefont
  {C.}~\bibnamefont {Renner}},\ }\bibfield  {title} {\bibinfo {title} {Scanning
  tunneling microscopy of the charge density wave in
  \textit{1T}-\textsc{T}i\textsc{S}e$_2$ in the presence of single atom
  defects},\ }\href@noop {} {\bibfield  {journal} {\bibinfo  {journal}
  {Physical Review B}\ }\textbf {\bibinfo {volume} {92}},\ \bibinfo {pages}
  {081101} (\bibinfo {year} {2015})}\BibitemShut {NoStop}%
\bibitem [{\citenamefont {Zunger}\ and\ \citenamefont
  {Freeman}(1978)}]{zunger1978band}%
  \BibitemOpen
  \bibfield  {author} {\bibinfo {author} {\bibfnamefont {A.}~\bibnamefont
  {Zunger}}\ and\ \bibinfo {author} {\bibfnamefont {A.~J.}\ \bibnamefont
  {Freeman}},\ }\bibfield  {title} {\bibinfo {title} {Band structure and
  lattice instability of \textsc{T}i\textsc{S}e$_2$},\ }\href@noop {}
  {\bibfield  {journal} {\bibinfo  {journal} {Physical Review B}\ }\textbf
  {\bibinfo {volume} {17}},\ \bibinfo {pages} {1839} (\bibinfo {year}
  {1978})}\BibitemShut {NoStop}%
\bibitem [{\citenamefont {Arguello}\ \emph {et~al.}(2014)\citenamefont
  {Arguello}, \citenamefont {Chockalingam}, \citenamefont {Rosenthal},
  \citenamefont {Zhao}, \citenamefont {Guti{\'e}rrez}, \citenamefont {Kang},
  \citenamefont {Chung}, \citenamefont {Fernandes}, \citenamefont {Jia},
  \citenamefont {Millis} \emph {et~al.}}]{arguello2014visualizing}%
  \BibitemOpen
  \bibfield  {author} {\bibinfo {author} {\bibfnamefont {C.~J.}\ \bibnamefont
  {Arguello}}, \bibinfo {author} {\bibfnamefont {S.~P.}\ \bibnamefont
  {Chockalingam}}, \bibinfo {author} {\bibfnamefont {E.~P.}\ \bibnamefont
  {Rosenthal}}, \bibinfo {author} {\bibfnamefont {L.}~\bibnamefont {Zhao}},
  \bibinfo {author} {\bibfnamefont {C.}~\bibnamefont {Guti{\'e}rrez}}, \bibinfo
  {author} {\bibfnamefont {J.}~\bibnamefont {Kang}}, \bibinfo {author}
  {\bibfnamefont {W.}~\bibnamefont {Chung}}, \bibinfo {author} {\bibfnamefont
  {R.~M.}\ \bibnamefont {Fernandes}}, \bibinfo {author} {\bibfnamefont
  {S.}~\bibnamefont {Jia}}, \bibinfo {author} {\bibfnamefont {A.~J.}\
  \bibnamefont {Millis}}, \emph {et~al.},\ }\bibfield  {title} {\bibinfo
  {title} {Visualizing the charge density wave transition in
  \textit{1H}-\textsc{N}b\textsc{S}e$_2$ in real space},\ }\href@noop {}
  {\bibfield  {journal} {\bibinfo  {journal} {Physical Review B}\ }\textbf
  {\bibinfo {volume} {89}},\ \bibinfo {pages} {235115} (\bibinfo {year}
  {2014})}\BibitemShut {NoStop}%
\end{thebibliography}%
\end{document}